%% file: main.tex
\documentclass[unnumsec,webpdf,contemporary,large]{style}%
\theoremstyle{thmstyleone}%
\theoremstyle{thmstyletwo}%
\theoremstyle{thmstylethree}%

\usepackage{booktabs, multirow} % for borders and merged ranges
\usepackage{soul}% for underlines
\usepackage{threeparttable} % for wide tables
\definecolor{myred}{RGB}{150,84,84}
\definecolor{myblue}{RGB}{0,47,167}
\usepackage{comment} 
\newcommand{\wf}[1]{\textcolor{red}{\bf\small [Fang WU: #1]}}
\newcommand{\xr}[1]{\textcolor{red}{\bf\small [Xiangru: #1]}}
\newcommand{\ek}[1]{\textcolor{blue}{\bf\small [elianna: #1]}}
\usepackage{chemformula}
\usepackage{titlesec}
\titleformat{\paragraph}[hang]{\normalfont\normalsize\bfseries}{\theparagraph}{1em}{}

\usepackage{wrapfig}
\usepackage{tcolorbox}
\usepackage{xcolor}
\usepackage{color}
\usepackage{colortbl}
\usepackage{adjustbox}

\newcommand{\mg}[1]{#1}
\newcommand{\mgg}[1]{#1}
\newcommand{\mggg}[1]{#1}
\newcommand{\bib}[1]{\textcolor{black}{#1}}
\newcommand{\bibb}[1]{\textcolor{black}{#1}}

\usepackage[edges]{forest}
\definecolor{hidden-draw}{RGB}{0,0,0}
\definecolor{hidden-pink}{RGB}{168,191,143}

\begin{document}
\appnotes{Published at Briefings in Bioinformatics}

\firstpage{1}

\title[New Frontiers in Molecule and Protein Generation]{A Survey of Generative AI for \textit{de novo} Drug Design: New Frontiers in Molecule and Protein Generation}

\author{Xiangru Tang$^{1, \ast \ORCID{0009-0006-2700-4513}}$} 
%$^{1,\ast}$
\author{Howard Dai$^{1, \ast}$}
\author{Elizabeth Knight$^{2, \ast}$}
\author{Fang Wu$^{3}$}
\author{Yunyang Li$^{1}$}
\author{Tianxiao Li$^{4}$}
\author{Mark Gerstein$^{1,4,5,6,7, \dagger \ORCID{0000-0002-9746-3719}}$}

\authormark{Tang et al.}

\address[1]{\orgname{Department of Computer Science}, \orgaddress{Yale University, New Haven, CT 06520}} 
\address[2]{\orgname{School of Medicine}, \orgaddress{Yale University, New Haven, CT 06520}} 
\address[3]{\orgname{Computer Science Department}, \orgaddress{Stanford University, CA 94305}}
\address[4]{\orgname{Program in Computational Biology \& Bioinformatics}, \orgaddress{Yale University, New Haven, CT 06520}}
\address[7]{\orgname{Department of Molecular Biophysics \& Biochemistry}, \orgaddress{Yale University, New Haven, CT 06520 
}}
\address[5]{\orgname{Department of Statistics \& Data Science}, \orgaddress{Yale University, New Haven, CT 06520 
}}
\address[6]{\orgname{Department of Biomedical Informatics \& Data Science}, \orgaddress{Yale University, New Haven, CT 06520}}

\corresp[$\ast$]{Contributed equally to this work.} 

\corresp[$\dagger$]{Corresponding author: Mark Gerstein. Email: mark@gersteinlab.org.}

\abstract{
Artificial intelligence (AI)-driven methods can vastly improve the historically costly drug design process, with various generative models already in widespread use. Generative models for \textit{de novo} drug design, in particular, focus on the creation of novel biological compounds entirely from scratch, representing a promising future direction. 
Rapid development in the field, combined with the inherent complexity of the drug design process, creates a difficult landscape for new researchers to enter. 
In this survey, we organize \textit{de novo} drug design into two overarching themes: small molecule and protein generation. Within each theme, we identify a variety of subtasks and applications, highlighting important datasets, benchmarks, and model architectures and comparing the performance of top models. 
We take a broad approach to AI-driven drug design, allowing for both micro-level comparisons of various methods within each subtask and macro-level observations across different fields. We discuss parallel challenges and approaches between the two applications and highlight future directions for AI-driven \textit{de novo} drug design as a whole. An organized repository of all covered sources is available at \url{https://github.com/gersteinlab/GenAI4Drug}. 
}


\maketitle

\section{Introduction}\label{intro}
During the drug design process, ligands must be created, selected, and tested for their interactions and chemical effects conditioned on specific targets \cite{drews2000drug}. These ligands range from small molecules with tens of atoms to large proteins such as monoclonal antibodies~\cite{mandal2009rational, colwell2018statistical}. While methods exist to optimize the selection and testing of probable molecules, traditional discovery methods across all fields are computationally expensive \cite{horvath2010comparison}. Recent artificial intelligence (AI) models have demonstrated competitive performance \cite{sliwoski2014computational, schneider2020rethinking, jing2018deep} in improving various tasks in the drug design process. Methods such as machine learning (ML)-driven quantitative structure-activity relationship (QSAR) approaches \cite{polishchuk2017interpretation, isarankura2009practical} have significantly improved virtual screening (VS) in molecule design \cite{li2017high, li2021machine}, while ML-assisted directed evolution techniques for protein engineering \cite{yang2019machine, wu2019machine} have proven to be reliable and widely used tools. However, an emerging and even more powerful task for ML is the generation of entirely new biological compounds in \textit{de novo} drug design \cite{hartenfeller2011novo, mouchlis2021advances, lima2016use}. 

\begin{figure*} 
\centering
\includegraphics[width=\textwidth]{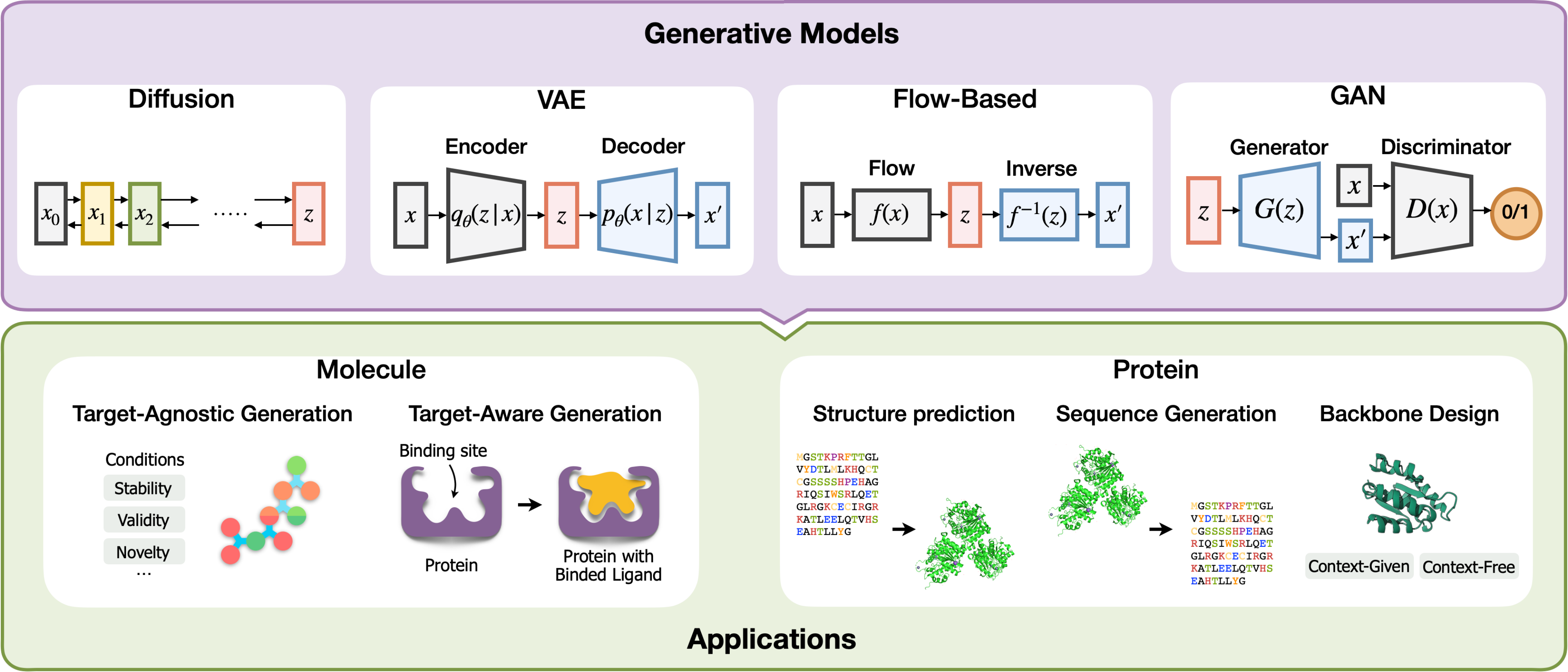}
\caption{An overview of the topics covered in this survey. In particular, we explore the intersection between generative AI model architectures and real-world applications, organized into two main categories: small molecule and protein generation tasks. \mggg{Note that diffusion and flow-based models are often paired with GNNs for processing 2D/3D-based input, while VAEs and GANs are typically used for 1D input \cite{dong2020molecular, Weng_2021, ganapathy2013crystal, Silva_2021, Protein, Embl-Ebi, zhao2018silico}.}
%Images used from \cite{dong2020molecular, Weng_2021, ganapathy2013crystal, Silva_2021, Protein, Embl-Ebi, zhao2018silico}.
}
\vspace{-.4cm}
\end{figure*}\label{fig:overview}

In contrast to applications in VS and directed evolution, which seek to expedite and optimize tasks within an existing framework, \textit{de novo} drug design focuses on generating entirely new biological entities not found in nature \cite{hartenfeller2011novo}. While other ML-driven methods search within existing chemical libraries for drug-like candidates, thereby facing inherent constraints, \textit{de novo} design circumvents this limitation by exploring unknown chemical space and generating drug-like candidates from scratch \cite{wang2022deep, kutchukian2010novo, liu2020computational}. 

%By definition, \textit{de novo} drug design is a generative task, and thus we focus on generative AI models. While predictive models are often used for loss functions and evaluative purposes, we do not discuss these models in detail. 

In this paper, we explore the impacts and developments of ML-driven \textit{de novo} drug design in two primary areas of research: \textbf{molecule} and \textbf{protein} generation. Within protein generation, we \bib{additionally explore antibody and peptide generation, given their} high research activity and relevance. Although the types of pharmaceuticals and the associated chemical nuances differ across fields, the overarching goal of exploring chemical space through \textit{de novo} design remains constant. \bib{Both} fields are rapidly growing industries with traditionally high research and development costs \cite{dimasi2015cost, lippow2007progress, zhou2016systems}, and current improvements are driven by active developments in ML-based \textit{de novo} methods.

\textit{Molecule design} specifically refers to the development of novel molecular compounds, often with the aim of small-molecule drug design. The generated molecules must satisfy a complex and often abstract array of chemical constraints that determine both their validity and ``drug-likeness" \cite{bickerton2012quantifying, ursu2011understanding}. This, combined with the vast space of potential drug-like compounds (up to $10^{23}$ - $10^{60}$ \cite{polishchuk2013estimation}), renders traditional small-drug design time-consuming and expensive. Using traditional methods, preclinical trials can cost hundreds of millions of dollars \cite{dimasi2016innovation} and take between 3 and 6 years \cite{horvath2010comparison}. In recent years, AI-driven methods have gained traction in drug design. AI-focused biotechnology companies have initiated over 150 small-molecule drugs in the discovery phase and 15 in clinical trials, with the usage of this AI-fueled process expanding by almost 40\% each year \cite{jayatunga2022ai}.

An equally promising field, \textit{protein design}, refers to the artificial generation or modification of proteins (protein engineering) for various biological uses. Native proteins have adapted and evolved over millions of years, so the rapid progression of human society in recent years poses challenges for naturally occurring proteins \cite{ding2022protein}. Protein design has an even more versatile range of applications, finding utility in immune signaling, targeted therapeutics, and various other fields. When executed efficiently, protein design has the potential to transform synthetic biology \cite{gao2020deep, huang2016coming, lippow2007progress}. Like molecule generation, proteins must adhere to abstract biological constraints, yet the inherently more complex structure of a protein presents a nuanced generative objective, requiring a more direct application of chemical knowledge in the process \cite{zhang2022ontoprotein, zhou2023protein, ma2023retrieved}. Traditional methods such as directed evolution are confined to specific evolutionary trajectories for existing proteins; the \textit{de novo} generation of proteins would add an entirely new dimension for researchers to explore \cite{romero2009exploring, huang2016coming, dahiyat1997novo}.

Structurally, we aim to provide an organized introduction to the two fields mentioned above. We begin with a technical overview of relevant deep learning architectures employed in both small molecule and protein design. We then explore their applications in molecule and protein design, dividing our analysis into a variety of subfields highlighted in Figure \ref{fig:overview}. For each subfield, we provide (1) a general background/task definition, (2) common datasets used for training and testing, (3) common evaluation metrics, (4) an overview of past and current ML approaches, and (5) a comparative analysis of the performance of state-of-the-art (SOTA) models. A detailed overview of this structure is shown in Figure \ref{fig:tree}. Finally, we integrate concepts within each subfield into a broad analysis of \textit{de novo} drug design as a whole, providing a comprehensive summary of the field in terms of current trends, top-performing models, and future directions. Our overall objective is to provide a systematic overview of ML in drug design, capturing recent advancements in this rapidly evolving area of research.

\vspace{-.1cm}

\section{Related Surveys}

Several survey papers delve into specific aspects of generative AI in drug design, with some focusing on molecule generation \cite{wang2022deep, zhang2023systematic, thomas2023integrating}, protein generation \cite{ding2022protein, gao2020deep}, or antibody generation \cite{akbar2022progress, hummer2022advances, chungyoun2023ai, kim2023computational}. Other survey papers are organized based on model architecture rather than application, with recent papers by Zhang et al. \cite{zhang2023survey} and Guo et al. \cite{guo2023diffusion} reviewing diffusion models in particular. While each of the above surveys provides an in-depth analysis of a specific application or type of model, the level of specialization may limit their scope. Our approach is a  macro-level analysis of small molecule and protein generation, tailored for those seeking a high-level introduction to the emerging field of generative AI in chemical innovation. This broad perspective enables us to highlight relationships across fields, such as parallel shifts in methods of input representation, the common emergence of architectures like equivariant graph neural networks (EGNNs), and similar challenges faced in both molecule and protein design.

\section{Preliminary: Generative AI Models}

Generative AI uses various statistical modeling, iterative training, and random sampling to generate new data samples resembling the input data. Historically, prominent approaches include generative adversarial networks (GANs) \cite{GANs2014}, variational autoencoders (VAEs) \cite{2013KingmaWelling}, and flow-based models \cite{rezende2015variational}. More recently, diffusion models \cite{Yang2023Yang} have emerged as promising alternatives. In our survey, we begin by providing a concise mathematical and computational overview of these architectures.
\vspace{-.2cm}

\subsection{Variational Autoencoders\href{https://medium.com/@j.zh/mathematics-behind-variational-autoencoders-c69297301957}{}}

% add history of AE to VAE here
A VAE \cite{2013KingmaWelling} is a type of generative model that extends upon the typical encoder-decoder framework by \mgg{representing each latent attribute using a distribution rather than a single value. This approach creates a more dynamic representation for the underlying properties of the training data and enables the sampling of new data points from scratch.} 
\clearpage
\input{./fig_tree}
\clearpage 
\noindent Formally, we can express the encoder as:
$$q_{\phi}(z|x) = \mathcal{N}(z;\mu_\phi(x),\sigma^2_\phi(x)I)$$
\mgg{Intuitively, each $x$ will be mapped to some mean $\mu_\phi(x)$ and variance $\sigma_\phi^2(x)$, which describe a corresponding normal distribution. We express the decoder as $p_\theta(x|z)$, where $z$  is a randomly sampled point from the latent distribution $\mathcal{N}(\mu_\phi(x),\sigma^2_\phi(x)I)$ and is mapped into a point $x$ in the decoding process. }

\mgg{VAE loss is computed using two balancing ideas: reconstruction loss and Kullback–Leibler (KL) divergence loss. Reconstruction loss measures the difference between the ground truth and the reconstructed decoder output, often expressed using cross-entropy loss:}
$$\mathcal{L}_{\text{recon}} = -1*\int{q_\phi(z|x)\log p_\theta(x|z) dz}$$
KL divergence measures the difference between two probability distributions \cite{van2014renyi}. \mgg{For VAEs, KL divergence is computed between the encoded distribution and the standard normal distribution. This can be seen as ``regularization," as it encourages the encoder to map elements to a more central region with overlapping distributions, thus improving continuity across the latent space. }
\bib{Formally, the KL loss can be expressed as follows, with $k$ representing the $k$th dimension in the latent space and $K$ representing the dimension of z:
$$\mathcal{L}_{\text{KL}} = D_{\text{KL}}(q_\phi(z|x^{(i)}) || \mathcal{N}(0, I))$$
$$= -\frac{1}{2} \sum_k \left(1 + \log({\sigma_k^{(i)}}^2) - {\mu_k^{(i)}}^2 - {\sigma_k^{(i)}}^2\right)$$ Here, $\mu_{k}^{(i)}$ and $\sigma_{k}^{(i)}$ represent the mean and variance of the kth component of the latent space, respectively, for datapoint $x^{(i)}$.} Then, the overall loss function can be expressed as follows, where $\beta$ can be adjusted to balance the reconstruction loss and KL loss: 
$$\mathcal{L} = \mathcal{L}_{\text{recon}} + \beta \mathcal{L}_{\text{KL}}$$

\subsection{Generative Adversarial Networks }

GANs \cite{GANs2014} utilize ``competing" neural networks for mutual improvement. The two neural networks --- the generator and the discriminator --- compete in a zero-sum game. The generator ($G$) creates instances (e.g., chemical structures of potential drugs) from random noise ($z$) sampled from a prior distribution $p_{z}(z)$ to mimic the training samples, while the discriminator ($D$) aims to distinguish between the synthetic data and the training samples.
% 
% 

%As this competition unfolds, the generator learns to produce more realistic outputs, and the discriminator becomes better at distinguishing real data from generated instances. The generator and discriminator are functions parameterized by $\theta_{g}$ and $\theta_{d}$ respectively. The discriminator outputs a single scalar $D(x;\theta_{d})$ representing the probability that $x$ comes from the real data, where $x$ can be real or generated data.
% 
% 

The learning process involves the optimization of the following loss function:
$$
\min_{G} \max_{D} \mathbb{E}_{x}[log D(x;\theta_{d})] + \mathbb{E}_{z \sim p(z)}[log(1-D(G(z;\theta_{g});\theta_{d}))]
$$
\mgg{Here, $\mathbb{E}_{x}[log D(x;\theta_{d})]$ represents the likelihood applied by the discriminator to a correct sample, while $\mathbb{E}_{z \sim p(z)}[log(1-D(G(z;\theta_{g});\theta_{d}))]$ represents the negative likelihood applied by the discriminator to an incorrect sample. This function returns a higher value when the discriminator accurately categorizes samples; thus, the discriminator aims to maximize this function, while the generator aims to minimize it. }

%The first term represents the expected value of the logarithm of the discriminator's output when given real data, and the second term represents the expected value of the logarithm of the discriminator being deceived by the generator's data.

\subsection{Flow-Based Models}
\mgg{Flow-based generative models \cite{rezende2015variational} generate data according to a target distribution $x \sim p(x)$ by applying a chain of transformations} to a simple latent distribution, often Gaussian, denoted $z_0 \sim p_0(z_0)$. This transformation applies an invertible function $f: z_0 \mapsto x$, such that: $$ x = f(z; \theta) \Rightarrow z = f^{-1}(x; \theta)$$
where the trained model learns parameters $\theta$. Since $f$ is invertible and thus the learned map is bijective,  $z$ has the same dimensionality as $x$. Often, $f$ is a composite function where $f(x) = f_K \circ f_{k-1} \circ ... \circ f_1(x)$; this allows for more complex probability distributions to be modeled. Because each function is invertible, \mgg{the posterior can be easily computed ---} the log-likelihood of a single point $x$ can be written in terms of its latent variable $z$: $$ \log p(x) = \log p_0(z) + \log \left| \text{det} \frac{\partial f}{\partial z} \right| $$ This function is used to train parameters $\theta$ to maximize the probability of observing the data. Various models build upon this premise to represent complex data distributions and capture relationships within sequential data. 

\subsection{Diffusion Models}

Diffusion models \cite{Yang2023Yang} perform a fixed learning procedure, \mgg{gradually adding Gaussian noise to data over a series of time steps. We define two stages of the model}: the noise-adding (forward) and the noise-removing (reverse) process. 

In the forward process, each step $x_{t+1}$ can be represented as a Markov chain and is comprised of $x_{t-1}$ and a small amount of Gaussian noise. We represent this mathematically as follows: 
\begin{align*} x_{t+1} = \sqrt{1-\beta_t}x_t + \sqrt{\beta_t}\epsilon, \quad \epsilon \sim \mathcal{N}(0, I) \end{align*}
Here, $x_t$ is the data at time t, and $\beta_t$ denotes the noise schedule. The variance $\beta_t$ decreases in the forward process so that after many steps, we have $p(x_t|x_0) \approx N(0,1)$. 

In the reverse process, our aim is to reconstruct the data from the noise. In this process, a denoising function is learned, often modeled by a neural network: 
\begin{align*}
x_{t-1} = f_{\theta}(x_t, t) + \sqrt{\beta_t}\epsilon, \quad \epsilon \sim \mathcal{N}(0, I)
\end{align*}

where $f_{\theta}(x_t, t)$ is the denoising function parameterized by $\theta$.

To train a diffusion model, we approximate the added noise at each step; the loss function minimizes the difference between the true noise and the model-predicted noise:
\begin{align*}
L_t = \mathbb{E}_{t \sim [1, T], x_0, \epsilon_t}[||\epsilon_t-\epsilon_\theta(x_t, t)||^2]
\end{align*}

Here, $t \sim U\{1, T\}$ means that the time step $t$ is drawn uniformly at random from the set $\{1,2,..., T\}$, and $\epsilon_{\theta}(x_t, t)$ represents the noise predicted by the model parameterized by $\theta$. $T$ represents the last time step of the model. Once the neural network has been trained, we can sample from the noise distribution and iterate through the reverse process to generate new data. 

% \mg{Diffusion models can be further categorized into three subtypes. \textit{Denoising diffusion probabilistic models (DDPMs)} \cite{ho2020denoising} focus on an iterative denoising process, where the reverse function is learned by a neural network that gradually removes the added noise. \mgg{\textit{Score-based generative models (SGMs)} \cite{song2019generative} instead estimate the gradient (score function) of the data distribution at each \mgg{timestep}, using it to subtract the error from the current representation. While the objective is framed differently, Ho et al. \cite{ho2020denoising} demonstrate that the SGM is equivalent to a DDPM during training, albeit under a different parameterization.} Finally, \textit{stochastic differential equation (SDE)-based models} \cite{song2020score} extend SGMs and DDPMs by learning continuous-time diffusion processes using Ito's calculus \cite{ikeda2012ito}. }

While the mathematical framework for the diffusion process is generally based on continuous data, adaptations made by Austin et al. \cite{austin2021structured} allow for smoother implementation with discrete data forms like molecular graphs. 

%These models leverage Ito's calculus to smoothly transform a complex data distribution into a known prior distribution by slowly injecting noise. A corresponding reverse-time SDE then finds the data distribution by solving for the score (time-dependent gradient field) of the perturbed data distribution; this score can be accurately estimated with neural networks. While both SDEs and SGMs use the gradient to model the error, SGMs sample via Langevin dynamics, whereas SDEs implement the score as part of a larger continuous differential equation. 

\subsection{Other Models}
\mggg{
While we cover the main generative methods used in these fields, a variety of other models also appear in specific applications and tasks in our paper, such as transformers, energy-based models (EBMs), BERT, and more \cite{vaswani2017attention, popescu2009multilayer, lecun2006tutorial, ngiam2011learning, schutt2017schnet, hochreiter1997long, devlin2018bert}. \bib{While not generative models on their own, graph neural networks (GNNs) \cite{scarselli2008graph} are often paired with the above generative methods to capture the graph-like structure of molecules. A wide variety of GNN variations exist, including equivariant graph neural networks (EGNNs) \cite{egnn}}, message-passing neural networks (MPNNs) \cite{gilmer2017neural}, graph convolutional networks (GCNs) \cite{kipf2016semi}, graph isomorphism networks (GINs) \cite{xu2018powerful}, and convolutional neural networks (CNNs) \cite{lecun1989backpropagation, gu2018recent, o2015introduction}. \bib{We discuss GNNs and EGNNs in more detail in the appendix on page \pageref{GNNs}.}}

%\subsection{Transformers}
%\mgg{
%A transformer model is a type of deep learning architecture first introduced in the context of natural language processing (NLP) that relies on attention mechanisms (self-attention) to enable the model to weigh the importance of different parts of the input data in relation to itself. This structure allows transformers to process sequential data, like text, efficiently. The self-attention mechanism relies on attention scores computed for each input sequence. The attention score uses three vectors (the query, key, and value vectors), which the model derives by applying transformations to the input based on the learned parameters. The dot product of the query vector with the key vector of every other element creates the attention score and communicates how much emphasis should be given to other parts of the input when processing the specific element. Models like AlphaFold, A-Prot, and ESMFold leverage the transformer architecture to predict a protein's three-dimensional structure from its amino acid sequence. Self-attention layers learn the relationships between amino acids and focus on the sequential proximity and interaction potential in the 3D folded structure. The ability to capture long-range dependencies within a sequence causes transformers to become a compelling tool for understanding protein sequences. }

\vspace{-.2cm}
\section{Applications}

\begin{comment}
%F

\end{comment}

\subsection{Molecule}
\begin{comment}
\subsubsection{Industry Background}
The drug design process has always been extremely costly. With traditional methods, preclinical trials cost hundreds of millions of dollars \cite{dimasi2016innovation} and take between 3 to 6 years \cite{horvath2010comparison}. In more recent years, AI-driven methods have found a place in drug design, as AI-focused biotech companies had over 150 small-molecule drugs in discovery and 15 in clinical trials, with the usage of this AI-fueled process expanding by almost 40\% each year \cite{jayatunga2022ai}. 
While AI is used for a multitude of functions such as property prediction, drug candidate ranking, and target identification, generative models focus specifically on the idea of \textit{\textit{de novo}} drug design. That is, molecular generative models aim to generate valid molecular compounds completely from scratch, circumventing the need for screening massive libraries of candidate molecules. This significantly reduces the need for brute-force trial and error, reducing both time and money costs for drug design. 
\end{comment}

\subsubsection{Task Background}
Molecule generation focuses on the creation of novel molecular compounds for drug design. These generated molecules are intended to be (1) valid, (2) stable, and (3) unique, with an overall goal of pharmaceutical applicability. ``Pharmaceutical applicability" is a broad term for a molecule's binding affinity to various biological targets. While the first three tasks may seem trivial, there are a variety of challenges with simply generating valid and stable molecules. Thus, the field of \textit{target-agnostic} molecule generation is focused on generating valid sets of molecules without consideration for any biological target. \textit{Target-aware} molecule generation (or ligand generation) focuses on the generation of molecules for specific protein structures and therefore focuses more on the pharmaceutical component. \mg{Finally, \textit{3D conformation generation} involves generating various 3D conformations given 2D connectivity graphs. }

For training and testing, molecule inputs can be formatted in a variety of ways, depending on the available information or desired output. Molecules can be expressed in 1D format through the simplified molecular-input line-entry system (SMILES), in 2D using connectivity graphs to represent bonds, or in 3D using point cloud embeddings on graph nodes~\cite{tang2023mollm}.

\begin{comment}
researchers in this field try to tackle the task of accommodating for a wide variety of restrictions and variables within any atom, considering a variety of interatomic relationships in the generation of complex and diverse molecules. Often, the molecules are represented as a fully-connected graph of nodes. Each node represents an atom, and has attributes including the 3-D coordinates and the element of the atom. The edges of the graph allows for the model to incorporate global information about the molecule as well as specific types of bonds between atoms in the molecule. One of the earliest papers used an equivariant diffusion model to generate molecules, where coordinates and features are generated by denoising from standard normal noise in \cite{hoogeboom2022equivariant}; while this model succeeded for small molecules, it had some difficulty scaling to larger molecules. Numerous other diffusion models incorporating other graph learning techniques worked in increasing the efficiency and scaling of the molecular generation. 

Papers in this field include: 
\cite{reward}
\cite{vignac2023midi}
\cite{morehead2023geometry}
\cite{huang2022mdm}
\cite{qiang2023coarse}
\cite{huang2023learning}
\cite{peng2023moldiff}
\cite{hua2023mudiff}
\cite{guan20233d}
\cite{arts2023two}
\cite{lin2022diffbp}
\cite{igashov2022equivariant}
\cite{wu2022diffusion}
\cite{bao2022equivariant}
\cite{wu2022score}
\cite{xu2022geodiff}
\cite{luo2021predicting}
\cite{jing2022torsional}
\cite{xu2023geometric}

\end{comment}

\subsubsection{Target-Agnostic Molecule Design}

\paragraph{Overview}
While the task of target-agnostic molecule design may seem simplistically open-ended, there are a vast array of chemical properties and rules that generated molecules must align with to be considered ``valid" and ``stable." The determination of ``validity" includes a complex combination of considerations, such as electromagnetic forces, energy levels, and geometric constraints, and a well-defined ``formula" does not yet exist for predicting the feasibility of molecular compounds. This, combined with the vast space of potential drug-like compounds (up to $10^{23}$ - $10^{60}$), makes brute-force experimentation quite time-consuming \cite{polishchuk2013estimation}. Deep learning can assist in learning abstract features for existing valid compounds and efficiently generate new molecules with a higher likelihood of validity. 

\begin{comment}
\xr{this paragraph is good, but please extend it and add more details or references}
\end{comment}

\paragraph{Task}
The target-agnostic molecule design task is as follows: given no input, generate a set of novel, valid, and stable molecules. 

\paragraph{Datasets} \label{GenMolDatasets}
To learn these abstract constraints, models must learn from large sets of existing valid, stable molecules. The following datasets are most commonly used for this task: 
\begin{itemize}
\item \textbf{QM9} \cite{ramakrishnan2014quantum}  - \textit{Quantum Machines 9}, contains small stable molecules pulled from the larger chemical universe database GDB-17
\item \textbf{GEOM-Drug}  \cite{axelrod2022geom} - \textit{Geometric Ensemble of Molecules}, contains more complex, drug-like molecules, often used to test scalability beyond the simpler molecules of QM9
\end{itemize}

\paragraph{Metrics}\label{Metrics:agnosticMolecule}
The most general task within this field is unconditional molecule generation, where models aim to generate a new set of valid, stable molecules with no input. All of these metrics can be evaluated using either QM9 or GEOM-Drug as testing sets. 

\begin{itemize}
    \item \textbf{Atom Stability} - The percentage of atoms with the correct valency
    \item \textbf{Molecule Stability} - The percentage of molecules whose atoms are all stable
    \item \textbf{Validity} - The percentage of stable molecules that are considered valid, often evaluated by RDKit
    \item \textbf{Uniqueness} - The percentage of valid molecules that are unique (not duplicates)
    \item \textbf{Novelty} - The percentage of molecules not contained within the training dataset
    \item \textbf{QED} \cite{bickerton2012quantifying} - \textit{Quantitative Estimate of Drug-Likeness}, a formulaic combination of a variety of molecular properties that collectively estimate how likely a molecule is to be used for pharmaceutical purposes 
    
\end{itemize}

Note that the novelty metric is sometimes omitted, as argued by Vignac et al. \cite{vignac2021top}, who contend that QM9 is an exhaustive set of all molecules with up to nine heavy atoms following a predefined set of constraints. Therefore, any ``novel" molecule would have to break one of these constraints, making novelty a poor indicator of performance. While QED is a well-established metric and may see expanded usage in the future, many current models have focused solely on generating valid molecules and do not report performance on QED. 

\begin{table*}[!htp]\centering
\scriptsize
\begin{tabular}{lccccccccc}\toprule
\textbf{Model} &\textbf{Type of Model} &\textbf{Dataset}  &\textbf{At Stb.} (\%, $\uparrow$) &\textbf{Mol Stb.} (\%, $\uparrow$)&\textbf{Valid} (\%, $\uparrow$)&\textbf{Val/Uniq.} (\%, $\uparrow$)\\\midrule
G-SchNet \cite{gebauer2019symmetry} &SchNet&QM9 &95.7 \cite{xu2023geometric}&68.1 \cite{xu2023geometric}&85.5 \cite{xu2023geometric}&80.3 \cite{xu2023geometric} \\
E-NF \cite{garcia2021n} &EGNN, Flow &QM9 &85 \cite{xu2023geometric}&4.9 \cite{xu2023geometric}&40.2 \cite{xu2023geometric}&39.4 \cite{xu2023geometric}\\
EDM \cite{xu2023geometric} &EGNN, Diffusion &QM9, GEOM-Drugs &98.7 &82.0 &91.9 &90.7 \\
GCDM \cite{morehead2023geometry}&EGNN, Diffusion &QM9 &98.7 &85.7 &94.8 &93.3 \\
MDM \cite{huang2022mdm} &EGNN, VAE, Diffusion &QM9, GEOM-Drugs & 99.2 \cite{huang2023learning} & 89.6 \cite{huang2023learning} &98.6 &94.6 \\
JODO \cite{huang2023learning} & EGNN, Diffusion & QM9, GEOM-Drugs & 99.2 & 93.4 & 99.0 & 96.0 \\
MiDi** \cite{vignac2023midi} &EGNN, Diffusion &QM9, GEOM-Drugs &99.8 &97.5 &97.9 &97.6 \\
GeoLDM** \cite{xu2023geometric} &VAE, Diffusion &QM9, GEOM-Drugs &98.9 &89.4 &93.8 &92.7 \\
\bottomrule
\end{tabular}
\vspace{.1cm}
\caption{An overview of relevant molecular generation models. All benchmarking metrics are self-reported unless otherwise noted. All metrics are evaluated with the QM9 dataset. For models with multiple variations, the highest-performing version was selected. [**] represents the current SOTA. As MiDi and MDM use slightly different evaluation conditions, their results are not fully comparable.}\label{tab:MoleculeQM9}
\vspace{-.4cm}
\end{table*}
\begin{table*}[!htp]
\centering
\scriptsize
\begin{tabular}{lccccc}\toprule
\textbf{Model} &\textbf{Atom Stb.} (\%, $\uparrow$) &\textbf{Mol Stb.} (\%, $\uparrow$) &\textbf{Valid} (\%, $\uparrow$) &\textbf{Val/Uniq.} (\%, $\uparrow$)\\\midrule
EDM \cite{hoogeboom2022equivariant} &81.30 & $\backslash$ &$\backslash$ &$\backslash$ \\
MDM \cite{huang2022mdm}&$\backslash$ &62.20 &99.50 &99.00 \\
MiDi \cite{vignac2023midi} &99.80 &91.60 &77.80 &77.80 \\
GeoLDM \cite{xu2023geometric} &84.40 & $\backslash$&99.30 &$\backslash$ \\
\bottomrule
\end{tabular}
\vspace{.1cm}
\caption{Molecular generation molecules evaluated on the larger GEOM-Drugs dataset. All metrics are self-reported. As MiDi uses slightly different evaluation conditions, its results are not fully comparable.  }\label{tab:MoleculeGEOM}

\end{table*}

\begin{table*}[!htp]
\centering
\scriptsize
\begin{tabular}{lrrrrrrr}\toprule
\textbf{Task ($\downarrow$)} &\textbf{$\boldsymbol{\alpha}$} &\textbf{$\boldsymbol{\Delta_{\varepsilon}}$} &\textbf{$\boldsymbol{\varepsilon_{HOMO}}$} &\textbf{$\boldsymbol{\varepsilon_{LUMO}}$} &\textbf{$\boldsymbol{\mu}$} &\textbf{$\boldsymbol{C_v}$} \\\cmidrule{1-7}
\textbf{Units} &$Bohr^3$ &$meV$ &$meV$ &$meV$ &$D$ &$\frac{cal}{mol}K$ \\\midrule
EDM \cite{hoogeboom2022equivariant} &2.76 &655 &356 &584 &1.111 &1.101 \\
GCDM \cite{morehead2023geometry} &1.97 &602 &344 &479 &0.844 &0.689 \\
MDM \cite{huang2022mdm} &1.591 &44 &19 &40 &1.177 &1.647 \\
GeoLDM \cite{xu2023geometric} &2.37 &587 &340 &522 &1.108 &1.025 \\
\bottomrule
\end{tabular}
\vspace{.1cm}
\caption{Molecular generation models evaluated on the conditional molecule generation task. All metrics are self-reported. \bib{All metrics are evaluated with the QM9 dataset. }}\label{tab:MoleculeConditional}
\vspace{-.4cm}

\end{table*}

Models are often evaluated on conditional molecule generation, aiming to generate models which fit desired chemical properties. For evaluation, a property classifier network $\phi_c$ is trained on half of the QM9 dataset, while the model is trained on the other half. $\phi_c$ is then evaluated on the model's generated molecules, and the mean absolute error between the target property value and the evaluated property value is calculated. Below are the six molecular properties considered: 
\begin{itemize}
\item \textbf{$\boldsymbol{\alpha}$ -} \textit{Polarizability}, or the tendency of a molecule to acquire an electric dipole moment when subjected to an external electric field, measured in cubic Bohr radius ($Bohr^3$)

\item \textbf{$\boldsymbol{\varepsilon_{HOMO}}$ -} \textit{Highest Occupied Molecular Orbit Energy}, measured in millielectron volts ($meV$)

\item \textbf{$\boldsymbol{\varepsilon_{LUMO}}$ -} \textit{Lowest Unoccupied Molecular Orbit Energy}, measured in millielectron volts ($meV$)

\item \textbf{$\boldsymbol{\Delta_{\varepsilon}}$ -} Difference between $\varepsilon_{HOMO}$ and $\varepsilon_{LUMO}$, measured in millielectron volts ($meV$)

\item \textbf{$\boldsymbol{\mu }$ -} \textit{Dipole moment}, measured in debyes ($D$)

\item \textbf{$\boldsymbol{C_v}$ -} \textit{Molar heat capacity} at 298.15 $K$, measured in calories per Kelvin per mole $\frac{cal}{mol}K$

\end{itemize}

\paragraph{Models}
Approaches to the molecular generation task have seen significant shifts over the past few years, transitioning from 1D SMILES strings to 2D connectivity graphs, then to 3D geometric structures, and finally to incorporating both 2D and 3D information.

Early methods like Character VAE (CVAE) \cite{gomez2018automatic}, Grammar VAE (GVAE) \cite{kusner2017grammar}, and Syntax-Directed VAE (SD-VAE) \cite{dai2018syntax} apply VAEs to 1D SMILES string representations of molecule graphs. While 1D SMILES strings can be deterministically mapped to molecular graphs, SMILES falls short in quality of representation -- two graphs with similar chemical structures may end up with very different SMILES strings, making it harder for models to learn these similarities and patterns \cite{jin2018junction}. 

Junction Tree VAE (JTVAE) \cite{jin2018junction} was the first model to address this issue by generating 2D graph structures directly. JTVAE generates a tree-structured scaffold and then converges this scaffold into a molecule using a graph message passing network. This approach allows JTVAE to iteratively expand its molecule and check for validity at each step, resulting in considerable performance improvement over previous SMILES-based methods. 

2D graph methods like JTVAE still fall short due to the lack of 3D input; because binding and interaction with other molecules/proteins rely heavily on 3D conformations, models that do not consider 3D information cannot properly represent and optimize properties like binding affinity. Thus, more recent developments include models that incorporate 3D information. Earlier 3D-based methods like E-NF \cite{garcia2021n} and G-SchNet \cite{gebauer2019symmetry} approached the molecular generation problem with \mgg{flow-based methods or autoregressive methods (in particular, G-SchNet uses the SchNet architecture developed by Schutt et al. \cite{schutt2017schnet})}. More recently, a wave of diffusion-based models operate on 3D point clouds, taking advantage of E(3) equivariance and demonstrating superior performance. 

EDM \cite{hoogeboom2022equivariant} provided an initial baseline for the application of diffusion, applying a standard diffusion process to an equivariant GNN with atoms represented as nodes with variables for both scalar features and 3D coordinates. While autoregressive models require an arbitrary ordering of the atoms, diffusion-based methods like EDM are not sequential and do not need such ordering, reducing a dimension of complexity and thus improving efficiency. 

Many subsequent models compared themselves with EDM as a baseline on the molecule generation task, seeking to improve upon its performance by adding additional considerations and adjustments. GCDM \cite{morehead2023geometry} implements a crossover between geometric deep learning and diffusion, using a geometry-complete perceptron network to introduce attention-based geometric message-passing. \mg{While both EDM and GCDM have already demonstrated massive performance improvements,} both models still struggle with both large-molecule scalability and diversity in the generated molecules. MDM \cite{huang2022mdm} addressed the scalability issue by pointing out the lack of consideration for interatomic relations in EDM and GCDM. MDM separately defines graph edges for covalent bonds and for Van der Waals forces (dependent on a physical distance threshold $\tau$) to allow for thorough consideration of interatomic forces and local constraints. In addition, MDM addressed the diversity issue by introducing an additional distribution-controlling noise variable in each diffusion step. While previous diffusion models operated directly in the complex atomic feature space, GeoLDM \cite{xu2023geometric} applies VAEs to map molecule structures to a lower-dimensional latent space for its diffusion. This latent space has a smoother distribution and lower dimensionality, leading to higher efficiency and scalability for large molecules. In addition, conditional generation is improved, as specified chemical properties are more clearly defined within latent spaces than they are in raw format. 

\begin{figure*}[t!]
\centering
\includegraphics[width = \textwidth]{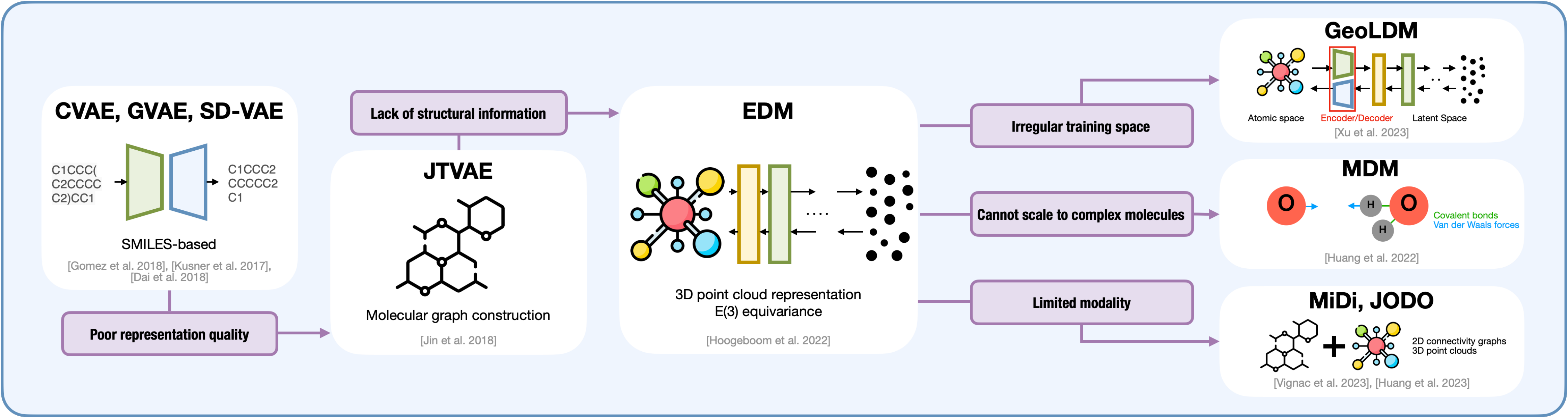}
\caption{An overview of the progress in target-agnostic molecule design over time. Shortcomings of previous models \bib{are shown in the corresponding pink boxes}, with subsequent models solving these shortcomings through novel design choices \cite{gomez2018automatic, hoogeboom2022equivariant, jin2018junction, xu2023geometric, huang2022mdm, huang2023learning}. 
%Images used from \cite{gomez2018automatic, hoogeboom2022equivariant, jin2018junction, xu2023geometric, huang2022mdm, huang2023learning}.
} 
\label{fig:mol}
\vspace{-.4cm}
\end{figure*}

While previous models learned exclusively from either 2D or 3D representations, a new wave of models recognize the need for both: a molecule's 2D connectivity structure is necessary to determine bond types and gather information about chemical properties and synthesis, while the 3D conformation is crucial for its interaction and binding affinity with other molecules. By jointly learning and generating both representations, models can maximize the amount of chemically relevant information and produce higher-quality molecular samples. The Joint 2D and 3D Diffusion Model (JODO) \cite{huang2023learning} uses a geometric graph representation to capture both 3D spatial information and connectivity information, applying score SDEs to this joint representation while proposing a diffusion graph transformer to parameterize the data prediction model and avoid the loss of correlation after noise is independently added to each separate channel. MiDi \cite{vignac2023midi} uses a similar graph representation but instead applies a DDPM. It proposes a ``relaxed" EGNN, which improves upon the classical EGNN architecture by exploiting the observation that translational invariance is not needed in the zero center-of-mass subspace. An full overview of the developments described in this section can be seen in Figure \ref{fig:mol}. 

As shown in Table \ref{tab:MoleculeQM9}, diffusion-based methods demonstrate significant improvements over previous methods, all achieving over 98.5\% in atom stability. However, some models fall behind when extended to the larger GEOM-Drugs dataset, as shown in Table \ref{tab:MoleculeGEOM}, where MiDi distinguishes itself for its capability to generate more stable complex molecules, albeit at the expense of validity. Table \ref{tab:MoleculeConditional} illustrates that MDM and GCDM excel at conditional generation tasks, with the former model achieving the best performance in four out of six tasks and the latter outperforming the remaining two. Overall, current models demonstrate high performance on the QM9 dataset, but there is room for improvement when dealing with the more complex molecules found in the GEOM-Drugs dataset.

\subsubsection{Target-Aware Molecule Design}
\paragraph{Overview}
Contrasting target-agnostic molecule design, target-aware design involves generating molecules based on specific biological targets. Within target-aware design, two primary approaches exist: ligand-based drug design (LBDD) and structure-based drug design (SBDD). LBDD models often utilize the amino acid sequences of target proteins, leveraging the characteristics and features of known ligands to build new molecules with similar properties. By contrast, SBDD models use the 3D structure of the target protein to design a corresponding molecular structure. LBDD models are most useful when the 3D structure is not experimentally available, but are limited in novelty because they only learn from existing bindings \cite{thomas2023integrating}. When the 3D structure of the target protein is available, SBDD models are generally preferred, as they consider crucial 3D information. 

\begin{comment} 
Tasks within target-aware molecule design include \textit{de novo} ligand generation, linker design, and fragment-based design. We describe and compare some models addressing these tasks.\\
\end{comment}
\paragraph{Task}
Given input target information, typically in the form of protein amino acid sequences in LBDD and protein 3D structure in SBDD, these approaches generate molecules that exhibit high binding affinity and potential interactions with this target. 
\paragraph{Datasets}
The following datasets are used for target-aware molecule design. CrossDocked2020 \cite{crossdocked2020} is currently the most heavily used, as the cross-docking technique allows for the generation of combinatorially large quantities of data by ``mixing and matching" similar ligand-protein pairs (22.5M compared to 40K in Binding MOAD \cite{hu2005binding}). 
\begin{itemize}
    \item \textbf{CrossDocked2020} \cite{crossdocked2020} - Contains ligand-protein complexes generated by cross-docking within clusters of similar binding sites called "pockets"
    \item \textbf{ZINC20} \cite{zinc20} - Fully enumerated dataset of possible ligands
\begin{comment}
    \item \textbf{CASF2} \cite{CASF2016} - Contains  experimentally verified 3D conformations and evaluation benchmarks
    \item \textbf{GEOM} \cite{axelrod2022geom} - Energy-annotated dataset of conformers 
\end{comment}
    \item \textbf{Binding MOAD} \cite{hu2005binding} - \textit{Binding Mother of All Databases}, a subset of PDB \cite{berman2000protein} containing experimentally determined protein-ligand pairings  
\end{itemize}

\paragraph{Metrics}
Target-aware molecule design utilizes the following metrics. Beyond affinity/applicability metrics, an additional consideration for diversity is considered, as a diverse array of potential options for a given target can provide more flexibility in the drug development process. 
\begin{itemize}
    \item \textbf{Vina Score} \cite{vina} - Scoring function supported by the Vina platform that returns a weighted sum of atomic interactions and is useful for docking
    \item \textbf{Vina Energy} \cite{vina} - Energy prediction by the Vina platform that is used to measure binding affinity
    \item \textbf{High Affinity Percentage} - Percentage of molecules with lower Vina energy than the reference (ground-truth) molecule when binding to the target protein
    \item \textbf{QED} \cite{bickerton2012quantifying} - \textit{Quantitative Estimate of Drug-Likeness}, which is also used in target-agnostic generation (see page \pageref{Metrics:agnosticMolecule})
    \item \textbf{SAscore} \cite{ertl2009estimation} - \textit{Synthetic Accessibility Score}, a formulaic combination of molecular properties that determine how easy a molecule is to create in a real lab setting
    \item \textbf{Diversity} - The diversity of generated molecules for each specific binding site, measured by Tanimoto similarities \cite{tanimoto1958elementary} between pairs of Morgan fingerprints 
\end{itemize}

\paragraph{Models}
LBDD models incorporate transformer architectures to generalize properties of learned ligands. For example, DrugGPT \cite{DrugGPT} is a recent autoregressive model that uses transformers to train numerous protein-ligand pairs. In their model, the ligand SMILES and protein amino acid sequence are tokenized for training, and the model produces viable SMILES ligand outputs. Generally, with improving protein structure prediction methods (see page \pageref{ProtStructurePrediction}) and increasing access to structural information as a whole, SBDD methods have become more prevalent than LBDD methods; thus, we explore SBDD methods below in more detail. 

LiGAN \cite{masuda2020generating} introduces the idea of a 3D target-aware molecule output, fitting molecules into grid formats for learning with a CNN and training the model under a VAE framework. Pocket2Mol \cite{peng2022pocket2mol} places more emphasis on the specific ``pockets" on the target protein to which the molecules bind, using an EGNN and geometric vector MLP layers to train on graph-structured input. Luo et al. \cite{luo20213d} directly model the probability of atoms occurring at a certain position in the binding site, taking advantage of invariance through the SchNet  \cite{schutt2017schnet} architecture.

Recent SBDD models have also popularized the use of diffusion models. 
\begin{comment}For example, DiffBP \cite{lin2022diffbp} uses a generative equivariant diffusion model that learns the conditional ligand distribution. Lin et al. also found that Geometric Vector Perceptrons (GVPs) are competitive with EGNN networks in learning transitional representations. 
\end{comment}
TargetDiff \cite{guan20233d} performs diffusion on an EGNN (with many similarities to EDM \cite{hoogeboom2022equivariant}) to learn the conditional distribution. To optimize the binding affinity, Guan et al. note that the flexibility of atom types should be low, which is reflected by the entropy of the atom embedding. Schneuing et al. \cite{schneuing2022structure} propose DiffSBDD, which includes two sub-models: DiffSBDD-cond and DiffSBDD-inpaint. DiffSBDD-cond is a conditional DDPM that learns a conditional distribution in a similar way as TargetDiff. In our benchmarking, we focus on the higher performing model, DiffSBDD-inpaint, which applies the inpainting approach (traditionally applied to filling parts of images) by masking and replacing segments of the ligand-protein complex.

\begin{table*}[!htp]
\centering
\scriptsize
\begin{tabular}{lcccccccc}\toprule
\textbf{Model} &\textbf{Type of Model} &\textbf{Dataset} &\textbf{Vina ($\uparrow$)} &\textbf{Affinity (\%, $\uparrow$)} &\textbf{QED ($\uparrow$)} &\textbf{SA ($\uparrow$)} &\textbf{Diversity ($\uparrow$)} \\\midrule
% DrugGPT \cite{DrugGPT} &Transformer &ZINC20 &\textbackslash &\textbackslash &\textbackslash &\textbackslash &\textbackslash \\
Luo et al. \cite{luo20213d} &SchNet &CrossDocked2020 &-6.344 &29.09 &0.525 &0.657 &0.720 \\
LiGAN \cite{masuda2020generating} &CNN, VAE &CrossDocked2020 &-6.144 \cite{luo20213d}&21.1 \cite{guan20233d} &0.39 \cite{guan20233d} &0.59 \cite{guan20233d} &0.66 \cite{guan20233d} \\
Pocket2Mol** \cite{peng2022pocket2mol} &EGNN, MLP &CrossDocked2020 &-5.14 \cite{guan20233d} &48.4 \cite{guan20233d} &0.56 \cite{guan20233d} &0.74 \cite{guan20233d} &0.69 \cite{guan20233d} \\
TargetDiff** \cite{guan20233d} &EGNN, Diffusion &CrossDocked2020 &-6.3 &58.1 &0.48 &0.58 &0.72 \\
DiffSBDD** \cite{schneuing2022structure} &EGNN, Diffusion &CrossDocked, MOAD &-7.333 &\textbackslash &0.467 &0.554 &0.758 \\
\bottomrule
\end{tabular}
\vspace{.1cm}
\scriptsize\caption{An overview of relevant target-aware molecular generation models. All benchmarking metrics are self-reported unless otherwise noted. [**] represents the current SOTA. As each paper uses slightly different benchmarking methods, their results may not be fully comparable. \bib{All metrics are evaluated with the CrossDocked2020 dataset. }}\label{tab:TargetAwareMolGen}
\vspace{-.4cm}
\end{table*}

As shown in Table \ref{tab:TargetAwareMolGen}, DiffSBDD leads in Vina score and diversity, while TargetDiff leads in high affinity. Interestingly, diffusion-based methods seem to be outperformed by the MLP used in Pocket2Mol when it comes to drug-like metrics like QED and SA. However, Guan et al. \cite{guan20233d} note that adjustments to TargetDiff, such as switching to fragment-based generation or predicting atom bonds, could improve performance on QED and SA. 

%TEMPORARY

\subsection{Protein}
\subsubsection{Task Background}
Proteins are large biomolecules that contain one or more long chains of amino acids. Each amino acid is a molecular compound that contains both an amino (-NH2) and a carboxylic acid (-COOH) \cite{lopez2020biochemistry}. While there are over 500 naturally occurring amino acids, only 22 are proteinogenic (``protein-building") and thus relevant to the protein generation task \cite{flissi2020norine}. Thus, in addition to 3D structural representations, proteins can be represented through their amino acid sequences, assigning each amino acid a letter label and representing each long chain as a string of labels. This sequential representation for amino acids mirrors the sequence structure of human language, allowing for natural language models to be applied in ways that would not be possible for the previously discussed molecule generation task. 
\mgg{Within protein generation, several generative subtasks can be defined. \textit{Representation learning} involves creating meaningful embeddings for protein inputs, which improves the data space for other models to train on. \textit{Structure prediction} involves the generation of a protein structure for its corresponding amino acid sequence, historically a challenging task due to the vast conformational space. \bib{\textit{Sequence generation} describes the inverse, creating a protein sequence for its corresponding structure.} Finally, \textit{\bib{backbone design}} refers to creating protein structures from scratch, which forms the core of the \textit{de novo} design task. }
% 
% For generating proteins, an efficient representation is needed, and several models focus on either sequence-based representation learning or structure-based representation learning. Many models follow the objective of generating the protein structure given the amino acid sequence. Some models focus on generating the backbone structure or the protein scaffold first. Then, they can apply other models focused on predicting the amino acid sequence or optimizing the side-chain packing of each residue given the scaffolds. Also, some models co-design the structure and sequence, which generalizes the applicable tasks. More specialized tasks for protein include working on protein-protein docking and protein-ligand interactions, which include possible conformation changes to the proteins in the complexes.

\mgg{We also briefly discuss antibody generation due to its high relevance within the protein generation field. In particular, antibodies are Y-shaped proteins used by the immune system to identify and bind to bacteria known as antigens. \mggg{While many protein design models use multiple sequence alignment (MSA) to map evolutionary relationships between related sequences, MSAs are not always available for antibody sequences, and antibody-specific models cannot rely on this input.} Additionally, binding regions are specifically defined for antibodies, contained within six complementarity-determining regions (CDRs). The CDR-H3 region is particularly diverse and complex, leading to a specialized task for the reconstruction of this region, known as CDR-H3 generation. \bib{We discuss antibody-specific methods within each corresponding subtask, and we include more detailed discussion of antibody generation in the appendix on page \pageref{antibodytaskbackground}.}}

\bib{Finally, we provide an additional section on peptide generation due to its high relevance and more specialized applications. Advances in drug delivery and synthesis technology have broadened its therapeutic potential, and recent innovations in peptide drug discovery have significantly improved treatments for type 2 diabetes with the creation of glucagon-like peptide-1 receptor agonists (e.g., liraglutide and semaglutide). Protein generation, while related, differs from peptide generation in length and complexity. Peptides are shorter (often no more than 50 amino acids) and have greater flexibility; thus, distinct computational models are needed to capture this differentiation.}

\subsubsection{Protein Representation Learning}
\paragraph{Overview}
\mgg{Protein representation learning involves learning embeddings to convert raw protein data into latent space representations, thereby extracting meaningful features and chemical attributes. Specifically, given a protein $x = [ o_1, o_2, \dots o_L ] $, where each $o_1$ represents an amino acid (sequence-based) or atom coordinate (structure-based), the goal is to learn an embedding $z = [h_1, h_2, \dots h_L]$, where each $h_i \in \mathbb{R}^d$ represents a $d-$dimensional token representation for amino acid $o_i$. Although representation learning is not a generative task on its own, these embeddings create ``richer" data spaces for other generative models to train on. Hence,  we briefly discuss them here. For a more in-depth analysis of protein representation learning models, please refer to the appendix on page \pageref{appendix}.}
\vspace{-0.1cm}
%TEMPORARY

\vspace{-0.05cm}
\vspace{-0.03cm}

\subsubsection{Structure Prediction}\label{ProtStructurePrediction}
\paragraph{Overview}
Generating 3D structures of proteins from their amino acid sequence is a challenging and important task in drug design. Historically, \bib{techniques like protein threading \cite{lemer1995protein} and homology modeling \cite{krieger2003homology} have been used to predict structures;} however, these methods have fallen short due to a lack of computational power and the difficulty of finding structures in the vast conformational space. New research in computational modeling has used various deep learning architectures to discover information from the amino acid sequences to generate accurate 3D structures. Current models have achieved impressive accuracy in structure prediction, but there is room for improvement in terms of speed and scale. \\
\vspace{-0.1cm}
\paragraph{Task}
Given a protein amino acid sequence, \bib{generate a set of 3D point coordinates for each amino acid residue, aiming to replicate a target ground-truth structure as closely as possible. }
\paragraph{Datasets} 
Unlike many of the other fields mentioned previously, the field of protein structure prediction benefits from a widely standardized benchmarking task through the Critical Assessment of Protein Structure Prediction (CASP). CASP conducts biennial testing on models using solved protein structures that have not been released to PDB.
\begin{itemize} \label{ProtStructurePredictionData}
\item \textbf{PDB} \cite{berman2000protein} - \textit{Protein Data Bank}, a central archive for all experimentally determined protein structures, widely used in almost all protein structure-related tasks
\item \textbf{CASP14} \cite{kryshtafovych2021critical} - \textit{14th Critical Assessment of Protein Structure Prediction}, a set of unreleased PDB structures used to create a standardized blind testing environment 
\item \textbf{CAMEO} \cite{haas2018continuous} - \textit{Continuous Automated Model Evaluation}, a complement to CASP that conducts weekly blind tests using around 20 pre-released PDB targets (to provide more continuous feedback, as CASP is biennial)

\end{itemize}

\paragraph{Metrics} 
Models are evaluated by comparing each protein's ground-truth structure with the generated structure. Three different approaches are taken to evaluate structural similarity:
\begin{itemize}
\item \textbf{RMSD} - \textit{Root-Mean-Square Deviation}, directly compares ground-truth positions with generated positions for each amino acid. If each $\delta_{i}$ denotes the distance between each ground truth and generated amino acid position, with $N$ total amino acids, we have: $$\mathrm {RMSD} ={\sqrt {{\frac {1}{N}}\sum _{i=1}^{N}\delta _{i}^{2}}}$$ 
\item \textbf{GDT-TS} \cite{zemla2003lga} - \textit{Global Distance Test-Total Score}, finds the most optimal superposition between two structures, searching for the highest number of corresponding residues that are within a distance threshold from each other. The GDT-TS aims  to represent global fold similarities.
\item \textbf{TM-score} \cite{zhang2004scoring} - \textit{Template Modeling score}, a similarity scoring formula that adjusts the GDT metric, normalizing for protein sequence length to avoid dependency on protein size, evaluating all residues beyond just those within the proposed cutoff for a more cohesive score. The TM-score aims to represent both global fold and local structural similarities.
\item \mgg{\textbf{LDDT} \cite{mariani2013lddt} - \textit{Local-Distance Difference Test}, a superpos.-free metric based on local distances between atoms. For each atom pair, its local distance is ``preserved" if the generated local distance is within a given threshold of the ground-truth distance, and the proportion of preserved distances is calculated. The LDDT can protect against artificially unfavorable scores when considering flexible proteins with multiple domains. }
 
\end{itemize}

\paragraph{Models}

\begin{table*}[!htp]\centering
\scriptsize
\begin{tabular}{lccccccc}\toprule
\multirow{2}{*}{\textbf{Model}} &\multirow{2}{*}{\textbf{Type of Model}} &\multirow{2}{*}{\textbf{Dataset}} & \multicolumn{4}{c}{\textbf{CAMEO}} & \multicolumn{1}{c}{\textbf{CASP14}} \\
& & & \textbf{RMSD (\r{A}, $\downarrow$)} &\textbf{TMScore ($\uparrow$)} & \textbf{GDT-TS} ($\uparrow$) & \mgg{\textbf{lDDT}} ($\uparrow$)& \textbf{TMScore ($\uparrow$)} 
\\\midrule
AlphaFold2** \cite{jumper2021highly}& Transformer & CASP14 & 3.30 \cite{jing2023eigenfold} & 0.87 \cite{jing2023eigenfold} & 0.86 \cite{jing2023eigenfold} &0.90 \cite{jing2023eigenfold} & 0.38 \cite{ESMFold} \\
RoseTTAFold \cite{Rosettafold}& Transformer & CAMEO, CASP14& 5.72 \cite{jing2023eigenfold} & 0.77 \cite{jing2023eigenfold} & 0.71 \cite{jing2023eigenfold} & 0.79 \cite{jing2023eigenfold}& 0.37 \cite{ESMFold}\\
ESMFold \cite{ESMFold}& Transformer & CAMEO, CASP14 & 3.99 \cite{jing2023eigenfold} & 0.85 \cite{jing2023eigenfold} & 0.83 \cite{jing2023eigenfold} & 0.87 \cite{jing2023eigenfold} &0.68 \\
EigenFold \cite{jing2023eigenfold}& Diffusion & CAMEO & 7.37 & 0.75 & 0.71 & 0.78 &  \textbackslash\\
\bottomrule
\end{tabular}
\vspace{.1cm}
\caption{An overview of relevant protein structure prediction models. All metrics are self-reported unless otherwise noted. Scores are provided as mean performance on single-structure tasks. lDDT represents the lDDT metric computed specifically using coordinates for the alpha carbon of each residue. [**] denotes the current SOTA.}\label{tab:ProteinStructurePredict}
\vspace{-.4cm}
\end{table*}

AlphaFold2 \cite{jumper2021highly} is a landmark model that uses deep learning techniques to compete with experimental methods. AlphaFold2 integrates numerous layers of transformers in an end-to-end approach. The transformers incorporate information from the MSA and pair representations to explore the folding space, potential orientations of amino acids, and overall structure based on pairwise distances. The MSA aligns multiple related protein sequences to create a 2D representation that informs the transformer architecture. Additionally, AlphaFold2 employs invariant point attention (IPA) for spatial attention, while the transformer captures interactions along the chain structure. Notably, AlphaFold2 introduces novel constraints from experimental data, which record probable distances between residues, preferred orientations between residues, and likely dihedral angles for the covalent bonds in the backbone.

Proposed in 2020, trRosetta \cite{trRosetta}, i.e., transform-restrained Rosetta, is another model that uses a deep residual network with attention mechanisms. Upon inputting MSA information, trRosetta predicts distances and orientations for residue pairs, which are then utilized to construct the 3D structure using a Rosetta protocol. Despite their advancements, both trRosetta and AlphaFold2 face several challenges, including their reliance on the MSA representation, limitations to natural proteins, and high computational requirements. Recently, RoseTTAFold \cite{Rosettafold}, which replaced trRosetta \cite{trRosetta}, demonstrated performance comparable to AlphaFold2 based on CASP14 test data. Importantly, RoseTTAFold can generate samples within 10 minutes, which is around 100 times faster than AlphaFold2. RoseTTAFold employs a three-track neural network that simultaneously learns from 1D sequence-level, 2D distance map-level, and 3D backbone coordinate-level information with attention mechanisms integrated throughout. RoseTTAFold exhibits robust performance in predicting protein complexes, whereas AlphaFold2 excels primarily for single protein structure prediction. \\

 Building on these techniques, ESMFold \cite{ESMFold} is a language model that predicts protein structure by leveraging ESM-2 representations. The output embeddings from ESM-2 are passed to a series of self-attending ``folding blocks." A structure module, featuring an SE(3) transformer architecture, generates the final structure predictions. Since ESMFold uses ESM-2 representations instead of MSA representations, this model offers faster processing with competitive scores on CAMEO and CASP14. EigenFold \cite{jing2023eigenfold} is another model that applies diffusion models to generate protein structures. The model represents the protein as a system of harmonic oscillators. Then, the structure can be projected onto the eigenmodes of that system during the forward process. In the reverse process, the rough global structure is sampled before refining local details. As a score-based model, EigenFold is not as computationally intensive but still underperforms in accuracy and range compared to other models. \\
\paragraph{Antibody Structure Prediction}
\mgg{A class of models has also been developed specifically catering to antibody structure prediction. As discussed previously, MSA alignment cannot be used for antibody input, rendering general models like AlphaFold highly inefficient and slow in the context of antibody prediction. IgFold \cite{ruffolo2023fast} uses sequence embeddings from AntiBERTy \cite{ruffolo2021deciphering} and invariant point attention to predict antibody structures, achieving SOTA generation speed with comparable accuracy to other models in the field. tFold-Ab \cite{wu2022tfold} performs comparably to IgFold, generating full-atom structures more efficiently by reducing reliance on external tools like Rosetta energy functions. For a more in-depth analysis of the datasets, task definition, metrics, and performance, refer to the appendix on page \pageref{AntibodyStructPredict}.}

\subsubsection{\bib{Sequence Generation}}
\paragraph{\bib{Overview}}
\bib{Sequence generation, also known as inverse folding or fixed-backbone design, entails the inverse task of structure prediction. Generating amino acid sequences that can fold into target structures is crucial for designing proteins with desired structural and functional properties. As with molecules and protein structures, the space of valid sequences is vast; this figure has been estimated to lie between $10^{65}$ and $10^{130}$ \cite{dryden2008much}. In addition, the process of protein folding is naturally complex and difficult to predict.}

\bib{To address these challenges, a variety of deep learning methods have been applied to represent the distribution of sequences with respect to structural information. While we briefly discuss some preliminary methods that are structure-agnostic, we place the highest focus on models that target specific protein structures. 
}

\paragraph{\bib{Task}}
\bib{Given a fixed protein backbone structure, generate a corresponding amino acid sequence that will fold into the given structure. }

\paragraph{\bib{Datasets}} \label{SequenceGenerationData}
\bib{Models in this field utilize the following datasets. Models primarily use CATH for training, with some using various augmentation methods utilizing UniRef and UniParc. CATH and TS500 are most frequently used for evaluation. To produce a standardized benchmark, Yu et al. \cite{yu2024multi} created a set of 14 known \textit{de novo} protein structures that do not exist in CATH to avoid data contamination. }
\begin{itemize}
\item \bib{\textbf{PDB} \cite{berman2000protein} - \textit{Protein Data Bank}, a comprehensive protein structure dataset (see page \pageref{ProtStructurePredictionData})}
\item\bib{\textbf{UniRef} \cite{apweiler2004uniprot} - A clustered version of the Unified Protein KnowledgeBase (UniProtKB), part of the central resource UniProt, which is a curated and labeled set of protein sequences and their functions}

\item\bib{\textbf{UniParc} \cite{apweiler2004uniprot} - A larger dataset of protein sequences, part of the central resource Uniprot, which includes UniProtKB and adds proteins from a variety of other sources}

\item\bib{\textbf{CATH} \cite{sillitoe2015cath} - A classification of protein domains (subsequences that can fold independently) into a hierarchical scheme with four levels: class (C), architecture (A), topology (T), and homologous superfamily (H). Using their classification, the authors also provide a diverse set of proteins that have minimal overlap and sequence similarity.}

\item\bib{\textbf{TS500} \cite{li2014direct} - A subset of 500 proteins from PDB \cite{wang2003pisces} filtered by sequence identity using the PISCES network. Li et al. also created a smaller subset, TS50, with additional filters to control for sequence length and fraction of surface residue.}
\end{itemize}

\paragraph{\bib{Metrics}}
\bib{While many models perform their own testing, we use results from the independent benchmark created by Yu et al. for fair comparison. We list the evaluated metrics below. Note that while Yu et al. did not evaluate on perplexity, we include it here due to its frequent use in sequence design method evaluations. }
\begin{itemize}
    \item \bib{\textbf{AAR} - \textit{Amino Acid Recovery}, the proportion of matching amino acids between the generated and native sequences}
    \item \bib{\textbf{Diversity} - The average difference between pairs of generated sequences, measured using Clustalw2 \cite{larkin2007clustal}}
    \item \bib{\textbf{RMSD} - \textit{Root-Mean-Square Deviation}, a structural comparison between two structures (see page \pageref{ProtStructurePredictionData}). In the context of sequence generation, the proposed sequences are folded into structures before comparison with native backbone structures.} 
    \item \bib{\textbf{Nonpolar Loss} - A metric measuring the rationality of polar amino acid types within the folded structure, where higher presence of nonpolar amino acids on the surface results in higher loss}
    \item \bib{\textbf{PPL -} \textit{Perplexity}, an exponentiation of cross-entropy loss, representing the inverse likelihood of a native sequence appearing in the predicted sequence distribution. For a series of $N$ amino acids $x_1, x_2, \dots x_N$, we can express perplexity as: 
$$
\textsc{PPL} = \text{exp}\left(\frac{1}{N} \sum_{i=1}^N \log P(x_i | x_{1}, x_2, \dots x_{i - 1})\right)$$}

\bib{Perplexity is calculated individually for each protein in the test set and averaged to produce a final PPL value. }
\end{itemize}

\paragraph{\bib{Models}}

\bib{A preliminary class of models generates novel protein sequences without considering a fixed backbone target, aiming to capture the unconditional distribution of amino acid sequence space.} ProteinVAE \cite{lyu2023proteinvae} utilizes ProtBERT \cite{elnaggar2021prottrans} to reduce raw input sequences into latent representations, employing an encoder-decoder framework with position-wise multi-head self-attention to capture long-range dependencies in these sequences. ProT-VAE \cite{sevgen2023prot} uses a different pre-trained language model ProtT5NV \cite{NVIDIA}. It includes an inner family-specific encoder-decoder layer to learn parameters relevant to specific protein families. Conversely, ProteinGAN \cite{repecka2021expanding} uses a GAN architecture to generate protein sequences. The model's efficacy is exemplified through the example of malate dehydrogenase, demonstrating its potential to generate fully functional enzyme proteins. \bib{While these approaches demonstrate relative success in generating valid and diverse sets of protein sequences, models that operate entirely in sequence space cannot consider crucial structural information. This limitation restricts their ability to capture the full range of constraints and dependencies between amino acid residues.} 

\bib{The primary class of models in this field receive fixed backbone targets as input, generating corresponding amino acid sequences. ProteinSolver \cite{strokach2020fast} draws connections between generating backbone structures and solving Sudoku puzzles, arguing that both are forms of constraint satisfaction problems (CSP) where positional information imposes constraints on the labels that can be assigned to each ``node." After finding a GNN architecture that can effectively solve Sudoku puzzles, Strokach et al. apply a similar architecture to the task of protein sequence design. In this design, node attributes encode amino acid types, and edge attributes encode relative distances between pairs of residues. PiFold \cite{gao2022pifold} extends this approach by introducing more comprehensive feature representations, including explicit distance, angle, and direction information in its node and edge features. Anand et al. \cite{anand2022protein} design a 3D CNN that directly learns conditional distributions for each residue given previous amino acid types, relative distances for heavy atoms, and torsional angles for side chains. Using these learned distributions, the 3D CNN autoregressively generates potential sequences. ABACUS-R \cite{liu2022rotamer} incorporates a pretrained transformer to infer a residue's amino acid types from nearby residues. To generate valid sequences, the model iteratively updates subsets of residues based on their environments, gradually constructing self-consistent sequences. ProRefiner \cite{zhou2023prorefiner} improves upon this design by introducing entropy scores for each prediction. While ABACUS-R uses every residue in the neighborhood for refinement, ProRefiner masks out high-entropy (low-confidence) predictions. By filtering out low-quality predictions, ProRefiner mitigates error accumulation from incorrect predictions. }
\begin{table*}[!htp]\centering

\scriptsize
\bib{\begin{tabular}{lcccccccc}\toprule
\textbf{Model} &\textbf{Type of Model} &\textbf{Dataset} &\textbf{AAR (\%, $\uparrow$)} &\textbf{Div.} ($\uparrow$) &\textbf{RMSD (\r{A}, $\downarrow$)} &\textbf{Non. ($\downarrow$)} &\textbf{Time (s, $\downarrow$)} \\\midrule
ProteinSolver \cite{strokach2020fast} &GNN &UniParc &24.6 &0.186 &5.354 &1.389 &180 \\
3D CNN \cite{anand2022protein} &CNN &CATH &44.5 &0.272 &1.62 &1.027 &536544 \\
ABACUS-R \cite{liu2022rotamer} &Transformer &CATH &45.7 &0.124 &1.482 &0.968 &233280 \\
PiFold \cite{gao2022pifold} &GNN &CATH, TS50/TS500 &42.8 &0.141 &1.592 &1.464 &221 \\
GVP-GNN \cite{jing2020learning} &GNN &CATH, TS50 &44.9* \cite{jing2020learning} &\textbackslash &\textbackslash &\textbackslash &\textbackslash \\
ESM-IF1** \cite{hsu2022learning} &Transformer &CATH, UniRef+ &47.7 &0.184 &1.265 &1.201 &1980 \\
ProteinMPNN** \cite{dauparas2022robust} &MPNN &CATH &48.7 &0.168 &1.019 &1.061 &112 \\
GPD \cite{mu2024graphormer} &Transformer &CATH &46.2 &0.219 &1.758 &1.333 & 35 \\
\bottomrule
\end{tabular}}
\vspace{.1cm}
\caption{ An overview of relevant sequence design methods. Div. and Non. refer to diversity and nonpolar loss, respectively. Results are reported by Yu et al. \cite{yu2024multi}. For GVP-GNN, we report self-evaluated AAR on a CATH test split. [**] denotes the current SOTA.}\label{tab:SequenceDesign}
\end{table*}

\bib{To better model input protein structures, GPD \cite{mu2024graphormer} uses the Graphormer \cite{ying2021transformers} architecture, which is a modified transformer for graph-structured data. GPD also uses Gaussian noise and random masks to improve diversity and recovery. GVP-GNN \cite{jing2020learning} uses a simple yet novel geometric representation for all nodes in the system. Rather than individually encoding vector features (like relative node orientations) for each node, these features are directly represented as geometric vectors that transform accordingly alongside global transformations. This approach defines a global geometric orientation rather than independent features of each node. ESM-IF1 \cite{hsu2022learning} extends upon the representations in GVP-GNN by attaching a generic transformer block and training on an expanded dataset. To generate additional training examples, ESM-IF1 uses MSA Transformer \cite{rao2021msa}, a representation learning model, to rank the sequences in the UniRef50 dataset by predicted LDDT scores. The top 12 million of these sequences are assigned corresponding structures through predictions made by AlphaFold2, producing a collection of sequence-structure pairs much larger than that of experimentally determined pairs. While previous methods require a fixed decoding order, ProteinMPNN \cite{dauparas2022robust} implements an order-agnostic autoregressive approach, which allows for a flexible choice of decoding order based on each specific task. }

\bib{We report the benchmark results measured by Yu et al. in Table \ref{tab:SequenceDesign}. ProteinMPNN generates the most accurate sequences, leading all methods in sequence recovery, RMSD, and nonpolar loss. GPD remains the most time-efficient method, generating sequences around three times faster than ProteinMPNN. Performance on diversity varies, but this can often be artificially controlled by adjusting a noise hyperparameter during testing to increase variation. Note that ProRefiner is not listed; ProRefiner primarily acts as an add-on module for existing methods and reports 2-7 percentage points of improvement on AAR when used to refine sequences for GVP-GNN, ESM-IF1, and ProteinMPNN.} 

\bib{In general, sequence generation remains a challenging field, as current SOTA models only recover fewer than half of the target amino acid residues.}

% \paragraph{Antibody Sequence Generation}
% Antibody 
% \vspace{-0.2cm}

\subsubsection{\bib{Backbone Design}}
\paragraph{Overview}
Like molecule generation, generating novel proteins from scratch can directly expand the library of available proteins capable of performing highly complex and versatile functions. While other areas such as structure prediction and \bib{sequence generation} contribute to the overall drug design process, \bib{backbone design} lies at the core of \textit{de novo} design, where new protein structures can be created entirely from scratch. 

\bib{As seen in molecule design, protein design contains} a similar distinction between structure and sequence. Some models generate 1D amino acid sequences, while others directly generate 3D structures, with some co-designing both 1D sequences and 3D structures.

\paragraph{Datasets}
Models in this field utilize the following datasets:
\begin{itemize}
    \item \textbf{PDB} \cite{berman2000protein} - \textit{Protein Data Bank}, a comprehensive protein structure dataset (see page \pageref{ProtStructurePredictionData})
    \item \textbf{AlphaFoldDB} \cite{varadi2022alphafold} - \textit{AlphaFold Database}, an expanded protein structural dataset created by using AlphaFold2 to predict structures for corresponding sequences in the UniRef dataset
    \item \textbf{SCOP} \cite{murzin1995scop} - \textit{Structural Classification of Proteins}, a classification of proteins by homology and structural similarity. SCOP has been updated several times to include additional categorizations and features, with many recent models using the extended SCOPe \cite{chandonia2022scope} database.
    \item \textbf{CATH} \cite{sillitoe2015cath} - A classification of protein domains into a hierarchical scheme with four levels, also used in sequence generation (see page \pageref{SequenceGenerationData})
\end{itemize}

\paragraph{Tasks}
The backbone design task involves designing a protein \bib{backbone structure} either from scratch or based on existing context. \bib{This involves generating coordinates} for the backbone atoms for each amino acid (nitrogen, alpha-carbon, carbonyl, and oxygen atom). External tools like Rosetta \cite{RosettaCommons} can be used for side-chain packing, generating the remaining atoms. 
\begin{itemize}
    \item \textbf{Context-Free Generation} - Given no input, the goal is to generate a \bib{diverse} set of protein structures. This task is evaluated using the self-consistency TM (scTM) score. 
    \item \textbf{Context-Given Generation} - This is an inpainting task for proteins. Given a motif (a set of existing amino acid residues for a native protein), the goal is to accurately fill in the missing residues according to the native protein, which is evaluated using a variety of similarity metrics like AAR, PPL, and RMSD.  
\end{itemize}

\paragraph{Metrics}
\bib{Generated backbones should be highly \textit{designable}.} High designability is generally determined by the ability of a structure to be generated by a corresponding amino acid sequence. While lab testing is optimal, this is not always feasible, so the folding process is simulated by other generative models. Thus, Trippe et al. \cite{trippe2022diffusion} proposed the scTM approach, which includes the following steps: a proposed structure is fed into a sequence-prediction model (typically ProteinMPNN \cite{dauparas2022robust}) to produce a corresponding amino acid sequence, and then fed back into a structure prediction model (typically AlphaFold2 \cite{jumper2021highly}) to produce a sample structure. The TM-score (see page \pageref{ProtStructurePrediction}) between the generated structure and this sample structure is calculated. 
\begin{itemize}
    \item \textbf{scTM} \cite{trippe2022diffusion} - \textit{Self-consistency TM-score}, an approach proposed by Trippe et al. to simulate the folding process, described in detail above. Scores of scTM $> 0.5$ are typically considered designable, so the percentage of generated structures with scTM $> 0.5$ is often reported. 
    \item \textbf{scRMSD} - \textit{Self-consistency RMSD}, identical to the scTM but uses RMSD instead of the TM-score for evaluation. A score of scRMSD $< 2$ is typically used as a cutoff.
    \item \textbf{AAR -} \textit{Amino acid recovery}, a comparison between the ground truth and the generated amino acid sequences. The AAR is also measured in antibody representation learning (see page \pageref{AntibodyRepLearning}).

    \item \textbf{RMSD} - \textit{Root-mean-square deviation}, measures distances between the ground truth and the generated residue coordinates. The RMSD is also used in protein structure prediction (see page \pageref{ProtStructurePrediction}).

\end{itemize}

\begin{comment}
Other models generate the scaffolding of proteins, mainly their backbone structure, and then predict the amino acid sequence. Each amino acid has a set of backbone atoms, consisting of a nitrogen atom, an alpha-carbon, a carbonyl carbon, and an oxygen atom. The amino acid can be described by the coordinates of these atoms, pairwise distances, and angles between relevant bonds. So, the models focus on generating the structure of the scaffold, and common architectures include GANS, VAEs, and diffusion models. To complete the proteins, they may append other models that predict the distributions of amino acids conditioned on the scaffolding structure.\\
\end{comment}

\paragraph{Models}

\begin{table*}[!htp]\centering
\scriptsize
\begin{tabular}{lcccccccc}\toprule
\multirow{2}{*}{\textbf{Model}} & \multirow{2}{*}{\textbf{Type of Model}} &\multirow{2}{*}{\textbf{Dataset}} & \multicolumn{2}{c}{\textbf{Context-Free}} & \multicolumn{3}{c}{\textbf{Context-Given}} \\ & & & \textbf{scTM (\%, $\uparrow$)} &\textbf{Design. (\%, $\uparrow$)} &\textbf{PPL ($\downarrow$)} &\textbf{AAR (\%, $\uparrow$)} &\textbf{RMSD (\r{A}, $\downarrow$)} \\\midrule
LatentDiff \cite{LatentDiff} &EGNN, Diffusion &PDB, AFDB &31.6 &\textbackslash &\textbackslash &\textbackslash &\textbackslash \\
FoldingDiff \cite{wu2022protein} &Diffusion &CATH &14.2 \cite{LatentDiff} &\textbackslash &\textbackslash &\textbackslash &\textbackslash \\
FrameDiff \cite{yim2023se} &Diffusion &PDB &84 &48.3 \cite{lin2023generating} &\textbackslash &\textbackslash &\textbackslash \\
Genie \cite{lin2023generating} &Diffusion &SCOPe, AFDB &81.5 &79.0 &\textbackslash &\textbackslash &\textbackslash \\
RFDiffusion** \cite{watson2023novo} &Diffusion &PDB &\textbackslash &95.1 \cite{lin2023generating} &\textbackslash &\textbackslash &\textbackslash \\
ProtDiff \cite{trippe2022diffusion} &EGNN, Diffusion &PDB &11.8 \cite{LatentDiff} &\textbackslash &\textbackslash &12.47* \cite{song2023joint} &8.01* \cite{song2023joint} \\
GeoPro \cite{song2023joint} &EGNN &PDB &\textbackslash &\textbackslash &\textbackslash &43.41* &2.98* \\
ProtSeed \cite{shi2022protein} &MLP &CATH &\textbackslash &\textbackslash &5.6 &43.8 &\textbackslash \\
%Anand et al. \cite{anand2022protein} &Diffusion &CATH &\textbackslash &\textbackslash &\textbackslash &\textbackslash &\textbackslash \\
Protpardelle \cite{chu2023all} &Diffusion &CATH &~85 &\textbackslash &\textbackslash &\textbackslash &\textbackslash \\
% GVP-GNN \cite{jing2020learning} &GVP-GNN &CATH &\textbackslash &\textbackslash &5.29 \cite{shi2022protein} &40.2 \cite{shi2022protein} &\textbackslash \\
\bottomrule
\end{tabular}
\vspace{.1cm}
\caption{An overview of relevant \bib{backbone design} methods. ``AFDB" refers to AlphaFoldDB. ``Design" refers to designability, defined by Lin et al. \cite{lin2023generating}, where any proteins with scRMSD $< 2$ and pLDDT $> 70$ (a local distance metric used in AlphaFold2 \cite{jumper2021highly}) are considered designable. \bib{All metrics are evaluated with the PDB  dataset, while }[*] denotes results tested only on $\beta$-lactamase metalloproteins extracted from PDB. [**] denotes the current SOTA.}\label{tab:ProteinDesign}
\vspace{-.4cm}
\end{table*}

% Another class of models focuses on generating novel protein structures. 
ProtDiff \cite{trippe2022diffusion} represents each residue with 3D Cartesian coordinates and uses a particle filtering diffusion approach. \mgg{However, 3D Cartesian point clouds do not mirror the folding process to create protein structures--- FoldingDiff \cite{wu2022protein} instead uses an angular representation for the protein structure, which more closely mirrors the rotational energy-optimizing protein folding process. FoldingDiff treats the protein backbone structure as a sequence of six angles representing orientations of consecutive residues. It denoises from a random, unfolded state to a folded structure using a DDPM and a BERT architecture.} LatentDiff \cite{LatentDiff} initially uses an equivariant protein autoencoder with GNNs to embed proteins into a latent space. Subsequently, it uses an equivariant diffusion model to learn the latent distribution. This process is analogous to GeoLDM \cite{xu2023geometric} for molecule design. Notably, LatentDiff's sampling on its latent space is ten times faster than sampling on raw protein space. 

The above models have shown relatively high performance in generating shorter proteins (up to 128 residues in length) but struggle with larger and more complex proteins \cite{lin2023generating}. To address longer protein structures, other methods use frame-based construction methods. This representation was initially demonstrated in the architecture of AlphaFold2 \cite{jumper2021highly} \mgg{in structure prediction, known as} IPA. In this paradigm, each residue is represented by orientation-preserving rigid body transformations (reference frames), which can be consistently defined regardless of global orientation. This allows for a more generalized representation than a series of 3D point clouds. Genie \cite{lin2023generating} performs discrete-time diffusion using a cloud of frames determined by a translation and rotation element to generate backbone structures. During each diffusion step, Genie computes the Frenet-Senet frames and uses paired residue representations and IPA for noise prediction. FrameDiff \cite{yim2023se} also parameterizes the backbone structures based on the frame manifold, using a score-based generative model. This approach establishes a diffusion process on $SE(3)^N$, the manifold of frames, invariant to translations and rotations. Then, the neural network predicts the denoised frame and torsion angle using IPA and a transformer model. Finally, RFDiffusion \cite{watson2023novo} combines the powerful structure prediction methods from RoseTTAFold with diffusion models to generate protein structures. RFDiffusion fine-tunes the RoseTTAFold weights and inputs a masked input sequence and random noise coordinates to iteratively generate the backbone structure. RFDiffusion also ``self-conditions" on the predicted final structure, leading to improved performance. RFDiffusion is a large pre-trained model with significantly more parameters than the other two frame-based models, enabling it to outperform the other frame-based models. \bib{GPDL \cite{zhang2023protein} utilizes a similar technique to RFDiffusion, using ESMFold instead of RoseTTAFold as its base structure prediction model. Additionally, it incorporates the ESM2 language model to extract important evolutionary information from input sequences to ESMFold. Due to ESMFold's superior efficiency in structure prediction, GPDL generates backbone structures 10-20 times faster than RFDiffusion.}

\bib{Another class of models aim} to co-design both protein sequence and structure simultaneously. 
GeoPro \cite{song2023joint} uses an EGNN to encode and predict 3D protein structures, designing a separate decoder to decode protein sequences from these learned representations. Protpardelle \cite{super} creates a ``superposition" over the possible sidechain states and collapses them during each iterative update step in the reverse diffusion process. The backbone is updated in each iterative step, while the sidechains are chosen probabilistically by another network to update. ProtSeed \cite{shi2022protein} uses a trigonometry-aware encoder that computes constraints and interactions from the context features and uses an equivariant decoder to translate proteins into their desired state, updating the sequence and structure in a one-shot manner. Anand et al. \cite{anand2022protein} use IPA as mentioned above, performing diffusion in frame space to efficiently generate protein sequences and structures. 

\begin{figure*}[t!]
\centering
\includegraphics[width = \textwidth ]{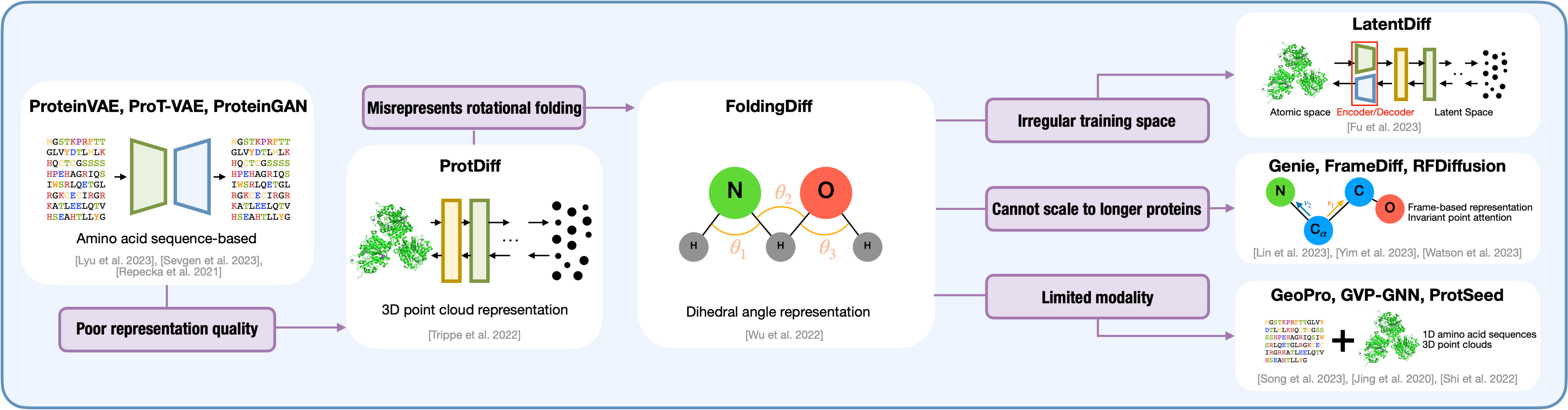}
\caption{\mgg{An overview of the progress in protein generation over time. Shortcomings of previous models \bib{are shown in the corresponding pink boxes}, with subsequent models solving these shortcomings through novel design choices \cite{sevgen2023prot, trippe2022diffusion, wu2022protein, LatentDiff, yim2023se, shi2022protein}. \bib{For consistency, only methods that generate proteins from scratch (without fixed backbone or sequence input) are depicted}. 
%Images used from \cite{sevgen2023prot, trippe2022diffusion, wu2022protein, LatentDiff, yim2023se, shi2022protein}.
}
}
\label{fig:prot}
\vspace{-.3cm}
\end{figure*}

For an overview of developments described in this section, see Figure \ref{fig:prot}. Note that as seen in molecule generation, we observe a progression from 1D-based (amino acid) models to 3D structure-based models to 1D/3D co-design; in addition, the field of protein design faces analogous questions of complexity scaling and latent space regularization. 

\paragraph{Antibody CDR-H3 Generation}
\mgg{As mentioned previously, antibody generation primarily focuses on the generation of a particular region known as the CDR-H3 region. Similar to protein generation, models in CDR-H3 generation have transitioned from sequence-based methods like the LSTM used by Akbar et al. \cite{akbar2022silico} to sequence-structure co-design pioneered by RefineGNN \cite{jin2021iterative} through iterative refinement. Notably, some models extend beyond the CDR-H3 generation task, aiming to tackle multiple parts of the antibody pipeline at once. dyMEAN \cite{kong2023end} is an end-to-end method incorporating structure prediction, docking, and CDR-H3 generation into a singular model. For a more in-depth analysis of datasets, task definition, metrics, and performance, refer to the appendix on page \pageref{AntibodyCDR}.}

\subsubsection{Peptide Design}
\paragraph{Overview}
% \bib{
% Peptides have high specificity, potent biological activity, and generally low toxicity, making them highly effective in targeting a wide array of proteins and thus disease states. Although peptides do have limitations, such as low oral bioavailability and rapid clearance from the bloodstream, advances in drug delivery and synthesis technology have broadened their therapeutic potential. Today, the peptide therapeutics market is robust, with around 80 drugs on the market and a significant number in development. For example, recent innovations in peptide drug discovery have significantly improved treatments for type 2 diabetes with the creation of the glucagon-like peptide-1 (GLP-1) receptor agonists (e.g. liraglutide, semaglutide).}

% Peptide generation is a crucial research area in computational biology and drug discovery, offering immense potential for personalized medicine and innovative treatments. Protein generation differs from peptide generation due to the inherent complexity and length of proteins compared to peptides. Peptides are shorter (often no more than 50 AAs) and have greater flexibility. Distinct computational models are needed to capture this differentiation. 

\bib{While we have discussed the monumental and robust models developed for protein generation, it is necessary to have models tailored for peptide-specific needs due to the inherently intricate and context-dependent nature of peptide structure, as well as the highly diverse array of downstream applications \cite{Muttenthaler2021}. This section briefly explores four different applications for AI in peptide generation, focusing on four state-of-the-art models: MMCD (Multi-Modal Contrastive Diffusion Model), PepGB (Peptide-protein interaction via Graph neural network for Binary prediction), PepHarmony, and AdaNovo.}

\paragraph{Peptide Generation}
\bib{In peptide generation, like protein backbone design, models aim to generate novel peptides from scratch. MMCD \cite{wang2024multimodal} is a diffusion-based model for therapeutic peptide generation that co-designs peptide sequences and structures (backbone coordinates). It employs a transformer encoder for sequences and an EGNN for structures, along with contrastive learning strategies to align sequence and structure embeddings and differentiate therapeutic and non-therapeutic peptide embeddings. MMCD outperforms baselines in both sequence and structure generation tasks, as demonstrated by testing on datasets of antimicrobial peptides and anticancer peptides. }
% In the antimicrobial peptides dataset (AMP), MMCD achieves a similarity score of 24.4107 (lower is better), instability score of 39.9649 (lower is better), and antimicrobial score of 0.8810 (higher is better), surpassing the best baseline model, SiamDiff, which obtains scores of 25.5385, 41.1629, and 0.8560. Similar performance improvements are observed in the anticancer dataset (ACP) and structure generation tasks. 

\paragraph{Peptide-Protein Interaction}
            \bib{For peptide-protein interactions, models aim to predict the physical binding site for a proposed peptide-protein pair. PepGB \cite{lei2024pepgb} is a GNN-based model for facilitating peptide drug discovery. It predicts peptide-protein interactions and leverages graph attention neural networks to learn interactions between peptides and proteins. PepGB was trained on a binary interaction benchmark dataset of protein-peptide and protein-protein interactions. A mutation dataset of peptide analogs targeting MDM2 was used for validating PepGB, and a large-scale peptide sequence dataset from UniProt was used for pre-training. PepGB consistently outperforms baselines in predicting peptide-protein interactions for novel peptides and proteins, with an increase of at least 9\%, 9\%, and 27\% in AUC-precision scores and 19\%, 6\%, and 4\% in AUC-recall scores under novel protein, peptide, and pair settings, respectively.  }
% It employs GAT as the backbone architecture (to update node features via aggregation form neighboring nodes), utilizing a fine-grained perturbation module to alleviate over-fitting, and optimizing binary cross-entropy loss and AUC min-max-margin loss to improve robustness and generalization.
\paragraph{Peptide Representation Learning}
\bib{ As with protein representation learning, models in peptide representation learning aim to convert raw peptide sequences into latent representations that capture valuable information. PepHarmony \cite{zhang2024pepharmony} is a multi-view contrastive learning model that integrates both sequence and structural information for enhanced peptide representation learning. It employs a sequence encoder (ESM) and a structure encoder (GearNet), which are trained together using contrastive or generative learning. PepHarmony utilizes data from conventional datasets like AlphaFoldDB and PDB while also employing a cell-penetrating peptide dataset (compiled from a variety of existing datasets), a solubility dataset (PROSOS-II \cite{smialowski2012proso}), and an affinity dataset (DrugBank \cite{wishart2008drugbank}). Zhang et al. report that PepHarmony demonstrates superior performance in downstream tasks such as cell-penetrating peptide prediction, peptide solubility prediction, peptide-protein affinity prediction, and self-contact map prediction. When compared to general protein representation learning methods like ESM2 and GearNet, PepHarmony outperforms baseline and fine-tuned versions of these models in most evaluation metrics, including accuracy, F1 score, area under the receiver operating characteristic curve, and correlation coefficients. }

% PepHarmonycl showed remarkable improvements, particularly in the CPP task, with an ACC of 0.79 and an F1 score of 0.766, which are the highest among the compared models. Furthermore, it achieved the highest ROC-AUC scores across all tasks.

\paragraph{Peptide Sequencing}
\bib{Mass spectrometry has played a crucial role in analyzing protein compositions from physical samples, but various forms of noise have posed challenges in extracting information from these reports. In peptide sequencing, models aim to address this challenge by predicting amino acid sequences given mass spectra data. AdaNovo \cite{xia2024adanovo} is a state-of-the-art model for \textit{de novo} peptide sequencing, composed of a mass spectrum encoder (MS Encoder) and two peptide decoders inspired by the transformer architecture. AdaNovo significantly improves upon previous models like DeepNovo \cite{tran2017novo}, PointNovo \cite{qiao2021computationally}, and Casanovo \cite{yilmaz2022novo} in terms of peptide-level and amino acid-level precision across various species. For example, in a human dataset, AdaNovo achieves a peptide-level precision of 0.373 and an amino acid-level precision of 0.618, outperforming DeepNovo (0.293 and 0.610), PointNovo (0.351 and 0.606), and Casanovo (0.343 and 0.585). AdaNovo's success is attributed to its innovative use of conditional mutual information and adaptive training strategies, which enhance its ability to identify post-translational modifications and handle noisy data typically associated with mass spectrometry.}

\vspace{-0.2cm}

\vspace{-.1cm}
\section{Current Trends}

\begin{comment}
\xr{Currently, you don't have the conclusions part in your draft. Summarize the paper, highlight what you believe are the best strategies, techniques or models, and suggest possible future work.}
- talk about fast improvements and movement in various fields, past original methods
% 
- talk about shift from purely sequence-based to structure-based to sequence/structure combined 
% 
- summarize most influential models/substantial advances in each section 
% 
- talk about directions for future developments
% 
\end{comment}

The drug design process, marked by a history of high complexity and cost, is poised for a transformative shift fueled by generative AI. AI-based methods are driving faster development and reducing costs, resulting in more effective and accessible pharmaceuticals for the public. Within the realm of generative AI, a notable shift has occurred; the emergence of GNNs and graph-based methods has fueled the transition from sequence-based approaches to structure-based approaches, ultimately leading to the integration of both sequence and structure in generation tasks. 
\begin{comment} Early sequence-based models like ProteinGAN(https://www.nature.com/articles/s42256-021-00310-5) and ProGen(https://pubmed.ncbi.nlm.nih.gov/36702895/) leverage GANs and transformer-based LLMs to predict novel protein sequences that code for an amino acid with particular physiochemical properties. Structure-based generative models use CNNs, VAEs, and equivariant graph networks to produce protein structures, like 3D coordinates or distance maps, to allow the incorporation of 3D geometric constraints. Pure structure models like FramDiff(https://arxiv.org/pdf/2302.02277.pdf) generate only 3D coordinates without associated sequences. Finally, sequence-structure models have been focused on during recent advancements and models like AlphaFold and RoseTTAFold jointly generate amino acid sequences and 3D structures in a coupled manner. 
\end{comment} 
% 
% 

Within the field of molecular generation, we are witnessing the recent dominance of graph-based diffusion models. These models take advantage of E(3) equivariance to achieve SOTA performance, with leaders like GeoLDM and MiDi excelling in target-agnostic design, and TargetDiff, Pocket2Mol, and DiffSBDD excelling in target-aware design. Finally, Torsional Diffusion outperforms all counterparts in molecular conformation generation. Additionally, we observe a shift from sequence to structural approaches in target-aware molecule design, where SBDD approaches demonstrate clear advantages over LBDD approaches, which operate with amino acid sequences. 
%Overall, while most models perform well generating stable molecules from QM9, we should see a shift towards a higher focus on larger molecules and drug applicability, moving towards testing on the more complex GEOM-Drugs dataset and comprehensive benchmarks like QED. 

%
% 

Within protein generation, a shift from sequence to structure is evident, as exemplified by the emergence of structure-based representation learning models like GearNET. These models leverage established sequence-based representation models such as ESM-1B and UniRep, recognizing the importance of 3D structural information in the protein generation process. AlphaFold2 remains a clear SOTA model for structure prediction. Similar to molecule generation, a wave of diffusion models are now tackling the protein scaffolding task, with RFDiffusion emerging as the top-performing model. 
% 
% 

% Finally, with antibody generation, we see a similar shift from sequence-based methods like LSTMs to sequence-structure co-design with methods like AntiDesigner and dyMEAN. Within each fragment of the complex antibody generation process, top models include PARA and AntiBERTa for representation learning, Igfold and tFold-Ab for structure prediction, EPMP for paratope/epitope prediction, and AntiDesigner for CDR generation. We also see a transition from single-task models to docking and CDR generation methods like DockGPT and, recently, a complete end-to-end method proposed with dyMEAN. 
\begin{comment}
Advanced transformer architectures and novel co-design strategies have empowered the task of protein generation. The astonishing accomplishments of AlphaFold began the excitement around the potential of foundational models to predict protein structure, and the tasks focus on structure prediction, protein-protein docking, and dynamics simulation. The best models and future directions…

Modern generative models for antibodies focus on the highly diverse CDR-H3 region and tackle assignments such as sequence learning, structure prediction, and antibody-antigen interactions, \ek{best models here} advances in simultaneous sequence and structure generation \ek{best models here}.

\end{comment}
\vspace{-0.3cm}

\section{Challenges and Future Directions}
While the prospects for generative AI in drug design are promising, several issues must be addressed before we can embrace ML-driven \textit{de novo} drug design. The main areas for improvement include increasing performance in existing tasks, defining more applicable tasks within molecule, protein, and antibody generation, and exploring entirely new areas of research. 

Current generative models still struggle with a variety of tasks and benchmarks. Within molecule generation, we face the following challenges:
\begin{itemize}
    \item \textbf{Complexity} - Models generate high frequencies of valid and stable molecules when trained on the simple QM9 dataset but struggle when trained on the more complex GEOM-Drugs dataset.
    \item \textbf{Applicability} - More applicable tasks like protein-molecule binding are especially challenging, and current models still struggle with generating molecules with high binding affinity for targets.
    \item \textbf{Explainability} - All methods discussed are fairly black-box and abstract; existing models do not reveal aspects like ``important" atoms or structures, and explainable AI in molecule generation is undeveloped as a whole.
\end{itemize}
Within protein generation, we encounter the following challenges:
\begin{itemize}
    \item \textbf{Benchmarking} - While most models in molecule generation use standardized benchmarking procedures, \textit{generative} tasks in protein design \bib{lack a standard evaluative procedure, with variance between each model's own metrics and testing conditions, making it hard to objectively evaluate the quality of designed proteins.} 
    \item \textbf{Performance} - As the tasks in protein generation are generally more \bib{complex} than those in molecule generation, \bib{SOTA models still struggle in several key areas like fold classification, gene ontology, and antibody CDR H3 generation, \bib{leaving room for future improvement.}} 
\end{itemize}

While our paper focuses on generative models and applications, it is important to note that many current tasks are evaluated with predictive models, such as the affinity optimization task in antibody generation or the conditional generation task for molecules. In these cases, classifier networks are used to predict binding affinity or molecular properties, and improvements to these classification methods would naturally lead to more precise alignment with real-world biological applications. 
% 
% 
\begin{comment}
Finally, the range of possibilities for generative AI in medicine is vast, and the generation of molecules, proteins, and antibodies constitutes only a primitive subset of what can be done. For example, tasks like RNA secondary structure prediction become increasingly important. RNA Secondary Structure can modulate transcription initiation, ribosomal movement, RNA stability, and splicing. The secondary structure of the RNA alters the accessibility of the ribosome binding site, and these secondary structures can slow ribosomes as they move along the mRNA during translation which can, in turn, modify the protein structure. Additionally, stable mRNAs allow for longer translation periods and potentially higher protein amounts. RNA secondary structures influence alternative splicing by hiding or exposing splice sites. The rise of antisense oligonucleotides (ASOs) as potential therapeutics for rare diseases accentuates the significance of RNA secondary structure in determining ASO viability in humans. ASO secondary structure directly affects mRNA binding affinity, target specificity, and stability, which, in turn, affects the efficacy of the therapeutic. Additionally, the secondary structure also modulates the immunogenicity of the ASOs, which often limits viability in vivo. Leveraging generative AI for understanding RNA secondary structure becomes vital to understanding RNA-driven cellular dynamics and to new RNA-based therapies.
\end{comment}
\vspace{-0.1cm}

\section{Conclusion}
%Our paper introduces a variety of current AI architectures, and then highlights exciting developments for generative AI in molecule and protein design. 

The survey has provided an overview of the current landscape of generative AI in \textit{de novo} drug design, focusing on molecule and protein generation. It has discussed important advancements in these fields, detailing the key datasets, model architectures, and evaluation metrics used. The paper also highlights key challenges and future directions, including improvements to benchmarking methods, improving explainability, and further alignment with real-world tasks to increase applicability. Overall, generative AI has shown great promise in the field of drug design, and continued research in this field can lead to exciting advancements in the future.

\section{Key Points}
\begin{itemize}

% \item A technical overview of primary model architectures in both small molecule and protein design.
% \item An outline of key questions within each field to highlight goals and define specific subtasks.
% \item A detailed collection of recent datasets, benchmarks, and generative models used within each subtask, including a comprehensive comparison of performance.
% \item A discussion of parallel developments and patterns across subtasks, providing a macro-level analysis between the fields of small molecule and protein design.
% \item A summary of remaining challenges and potential research directions, characterizing the future trajectory for AI-driven drug design.
    \item Our survey examines the advancements and applications of Generative AI within \textit{de novo} drug design, particularly focusing on the generation of novel small molecules and proteins.

    \item We explore the intricacies of generating biologically plausible and pharmaceutically potential compounds from scratch, providing a comprehensive yet approachable digest of formal task definitions, datasets, benchmarks, and model types in each field. 

    \item The work captures the progression of AI model architectures in drug design, highlighting the emergence of equivariant graph neural networks and diffusion models as key drivers in recent work. 

    \item We highlight remaining challenges in applicability, performance, and scalability, delineating future research trajectories. 
%    \item Maintaining a balance between technical depth and accessibility, this survey provides readers with a comprehensive yet approachable digest of current trends and methodologies.
    
    \item Through our organized repository, we aim to facilitate further collaboration in the rapidly evolving intersection of computational biology and artificial intelligence. 

\end{itemize}

\section{Funding}

Xiangru Tang and Mark Gerstein are supported by Schmidt Futures.

\bibliographystyle{unsrt}
\bibliography{main}

%USE THE BELOW OPTIONS IN CASE YOU NEED AUTHOR YEAR FORMAT.
%\bibliographystyle{abbrvnat}
%\bibliography{reference}

\clearpage

\section{Appendix} \label{appendix}

\subsection{Graph Neural Networks} \label{GNNs}
\bib{While graph neural networks (GNNs) \cite{scarselli2008graph} are \mgg{not inherently generative models on their own, they represent important components of larger generative methods}. \mgg{GNNs are especially important for processing graph-structured data. In the context of proteins, nodes and edges represent amino acids and spatial or sequential proximity, respectively. Once data have been transformed into a graph \( G = (V, E) \) with nodes \( V \) and edges \( E \), the GNN \bibb{$\phi$} learns to map nodes to embeddings through message passing (\bibb{$\phi_e$}) and aggregation (\bibb{$\phi_h$}). For each pair of nodes \( v_i, v_j \), the GNN \bibb{outputs} a ``message" $\mathbf{m}_{ij}$ based on existing features $\mathbf{h}_i^l, \mathbf{h}_j^l$ \mggg{and coordinates $\mathbf{x}_i^l, \mathbf{x}_j^l$} at layer $l$ \mggg{($a_{ij}$ denotes the $(i, j)$ entry in adjacency matrix $A$)}:
\mggg{$$\mathbf{m}_{ij} = \phi_e(\mathbf{h}_i^l, \mathbf{h}_j^l, \mathbf{x}_i^l, \mathbf{x}_j^l, a_{ij})$$}
Then, the message received by any node $v_i$ is an aggregation of the messages from its neighbors:
$$\mathbf{m}_i = \sum_{j\in \mathcal{N}(i)} \mathbf{m}_{ij}$$
Finally, \bibb{the model combines} old embeddings \mggg{and positions} with messages to create new embeddings \mggg{and positions} at layer $l + 1$:
\mggg{$$\mathbf{h}_i^{l+1}, \mathbf{x}_i^{l+1} = \phi_h(\mathbf{h}_i^l, \mathbf{x}_i^l, \mathbf{m}_i)$$}}
\mggg{Note that positions $\mathbf{x}$ and features $\mathbf{h}$ are often concatenated and denoted as a singular feature vector $\mathbf{h}$, as they are treated equivalently in the general GNN case.}}

% \[
% h_v^{(k)} = \sigma \left( \sum_{j\in \text{Neighbors}}{W_k\cdot h_j^{(k-1)}}\right) 
% \]

% \( h_v^{(k)} \) is the feature of node \( v \) at layer \( k \) and \( \sigma \) is a non-linear activation function, and \( W_k \) is a learnable weight matrix for layer \( k \). \mgg{The summation represents an arbitrary aggregation function}, but different neural networks can customize this choice. 
% \mgg{For graph-level tasks, node features} are combined with the $\text{AGG}$ function and create a graph-level output $\text{OUT}(G) = \text{AGG} \left( \{ h_v^{(K)} : v \in V \} \right)$. 

\subsubsection{Equivariant Graph Neural Networks }
\begin{comment}
    Equivariant Graph Neural Networks (EGNNs) inherently weave symmetry's inductive bias into model architectures. Traditional EGNNs predominantly lean on group theory, invoking irreducible representations~\cite{thomas2018tensor, fuchs2020se, anderson2019cormorant} and leveraging the Clebsch-Gordan (CG) product to guarantee equivariance. Contemporary strides in this domain, such as those presented in~\cite{batzner20223, frank2022sokrates, batatia2022mace}, have incorporated higher-order spherical harmonic tensors, resulting in notable performance enhancements. Parallel to these, alternative advancements have been steered by methods like PaiNN~\cite{schutt2021equivariant} and TorchMD-NET~\cite{tholke2022torchmd}, which prioritize iterative updates of scalar and vectorial features. ViSNet~\cite{wang2022visnet} builds upon PaiNN's foundation, integrating runtime geometric computation (RGC) and vector-scalar interactive message passing (ViS-MP). 
has proposed a fragmentation-based framework.
\end{comment}
\bib{\mg{For the generation of 3D structures, equivariance is a useful inductive bias to incorporate into deep learning models.} \mgg{In general, two 3D structures should be treated equally if they only differ under a series of rotations, reflections, and translations. }}

\bib{\mgg{For some conditional distribution $p(y|x)$, it is equivariant to the action of rotations and reflections when $p(Ry) = Rp(y)$ (alternatively, $p(y|x) = p(Ry|Rx)$) for transformation $R$. A distribution is invariant if  $p(Ry) = p(y)$. In most works, the transformations consist of the Euclidean E(3) group generated by rotations, translations, and reflections. }}

\bib{\mgg{Satorras et al. \cite{egnn} propose the EGNN, a simple adjustment to the traditional GNN framework to preserve equivariance. This is accomplished by defining separate point embeddings $\mathbf{x}_i$ and feature embeddings $\mathbf{h}_i$ for each node $v_i$, and considering relative positions for message passing:
$$\mathbf{m}_{ij} = \phi_e(\mathbf{h}_i^l, \mathbf{h}_j^l, \| \mathbf{x}_i^l - \mathbf{x}_j^l  \|, a_{ij})$$
$$\mathbf{x}_i^{l+1} = \mathbf{x}_i^l + C \sum_{j\neq i} (\mathbf{x}_i^l - \mathbf{x}_j^l) \phi_x(\mathbf{m}_{ij})$$
The final message aggregating and embedding update steps are equivalent to those of a GNN. Here, equivariance is preserved because we only consider relative positions/distances $\mathbf{x}_i^l - \mathbf{x}_j^l$ in message passing. Therefore, rotating, translating, or reflecting all atoms will result in equivalent functions. }}

\subsubsection{Molecular Conformation Generation}
\paragraph{Overview}
While learning on graph-structured data has found success in molecular generation, recent models have demonstrated that incorporating 3D information is crucial to understanding various chemical properties such as binding affinity. However, deriving a 3D structure from a 2D connectivity graph is nontrivial; given a 2D graph, there are numerous ways to construct a 3D arrangement of molecules that still adheres to the same connectivity structure. Two molecules with the same chemical formula but different 3D arrangements are called conformers (or conformational isomers). 

\paragraph{Datasets}
The following datasets are used for conformation generation. Note that while GEOM-QM9 \cite{axelrod2022geom} is created with the same molecules as in the general QM9 \cite{ramakrishnan2014quantum} dataset, the GEOM-QM9 dataset contains a complete set of representative conformers for each molecule.  QM9 only provides singular 3D structures and cannot be applied to conformation generation. 
\begin{itemize}
    % \item \textbf{COD} \cite{gravzulis2009crystallography} - \textit{Crystallography Open Database}, a collection containing all inorganic, metal-organic, and small organic structural data. Reference conformations are contributed voluntarily and can come from different sources/environments.
    \item \textbf{GEOM-QM9} \cite{axelrod2022geom} - Set of all conformers for each molecule in QM9. Reference conformations are created through a standardized process using density functional theory.
    \item \textbf{GEOM-Drugs} \cite{axelrod2022geom} - Conformers for more complex drug-like molecules, also used in general molecule generation (see page \pageref{GenMolDatasets})
    \item \textbf{ISO17} \cite{schutt2017schnet} - Set of molecular conformations for all molecules with the chemical formula \ch{C7O2H10}

\end{itemize}

\paragraph{Task}
Given a 2D molecule connectivity graph, this approach generates a set of stable 3D conformations that correspond to the given connectivity structure. 

\paragraph{Metrics}
While other tasks like property prediction are often evaluated on models in this field, we focus on generative tasks like conformation generation. The following metrics are used to judge the quality of the conformation generation task. Intuitively, COV measures the diversity of the generated sample, while MAT measures quality/accuracy. 
\begin{itemize}
    \item \textbf{COV} \cite{xu2021learning} - \textit{Coverage}, a percentage of how many ground-truth conformations are ``covered" (below a threshold root-mean-square deviation (RMSD) with some generated conformation, typically 0.5Å for QM9 and 1.25Å for Drugs).
    \item \textbf{MAT} \cite{xu2021learning} - \textit{Matching}, the average RMSD between each generated conformation and the closest ground-truth conformation.
\end{itemize}

\paragraph{Models}
Mansimov et al. propose a conditional variational graph autoencoder (CVGAE) \cite{mansimov2019molecular} for molecule generation, directly applying a GNN to learn atom representations and generate 3D coordinates. GraphDG \cite{simm2019generative} improves upon this by modeling distance geometry between atoms, which incorporates rotational and translational invariance (i.e., two identical molecules are treated equally under transformation). Conditional graph continuous flow (CGCF) \cite{xu2021learning} applies a flow-based approach to model the distance geometry, combining this with an (EBM) as a ``tilting" term for mutual enhancement. GeoMol \cite{ganea2021geomol} argues that methods solving for distance geometries are flawed, as creating distance matrices leads to overparameterization, and important geometric features like torsional angles are not directly modeled. To address this, GeoMol uses an MPNN to individually predict the local structure for each atom, allowing for a more direct representation of neighboring features. ConfGF \cite{shi2021learning} pulls ideas from force field methods in molecular dynamics, directly learning gradient fields of the log density for atom coordinates, using a GIN \cite{hu2021forcenet} to adjust for rotational and translational invariance. Dynamic graph score matching (DGSM) \cite{luo2021predicting} extends the ideas of ConfGF by constructing dynamic graphs to incorporate long-range atomic interactions (instead of relying on static input graphs that only consider chemical bonds). GeoDiff \cite{xu2022geodiff} uses an Euclidean space diffusion process, treating each atom as a particle. GeoDiff incorporates Markov kernels that preserve equivariance, as well as roto-translational invariance \mggg{through their proposed graph field network (GFN) layer, which draws inspiration from the eGNN architecture \cite{egnn}}. Torsional Diffusion \cite{jing2022torsional} uses score-based diffusion in the space of torsion angles, which allows for improved representation and fewer denoising steps than GeoDiff. 

\begin{table*}[!htp]\centering
\begin{tabular}{lcccccc}\toprule
\multirow{2}{*}{\textbf{Model}} & \multirow{2}{*}{\textbf{Type of Model}} &\multirow{2}{*}{\textbf{Dataset}} & \multicolumn{2}{c}{\textbf{GEOM-QM9}} & \multicolumn{2}{c}{\textbf{GEOM-Drugs}} \\
& & & \textbf{COV (\%, $\uparrow$)}& \textbf{MAT (\r{A}, $\downarrow$)}&  \textbf{COV (\%, $\uparrow$)}& \textbf{MAT (\r{A}, $\downarrow$)} \\\midrule
CVGAE \cite{mansimov2019molecular}&VAE, MPNN &QM9, COD, CSD &0.09 \cite{luo2021predicting} &1.6713 \cite{luo2021predicting} &0.00 \cite{luo2021predicting} &3.0702 \cite{luo2021predicting} \\
GraphDG \cite{simm2019generative}&VAE &ISO17 &73.33 \cite{luo2021predicting} &0.4245 \cite{luo2021predicting} &8.27 \cite{luo2021predicting} &1.9722 \cite{luo2021predicting} \\
CGCF \cite{xu2021learning} &Flow, EBM &G-QM9, G-Drugs, ISO17 &78.05 \cite{xu2022geodiff} &0.4219 \cite{xu2022geodiff} &53.96 \cite{xu2022geodiff} &1.2487 \cite{xu2022geodiff} \\
ConfGF \cite{shi2021learning}&GIN, Diffusion &G-QM9, G-Drugs &88.49 \cite{luo2021predicting} &0.2673 \cite{luo2021predicting} &62.15 \cite{luo2021predicting} &1.1629 \cite{luo2021predicting} \\
GeoDiff** \cite{xu2022geodiff}&GFN, Diffusion &G-QM9, G-Drugs &90.07 &0.209 &89.13 &0.8629 \\
GeoMol \cite{ganea2021geomol}&MPNN &G-QM9, G-Drugs &71.26 \cite{xu2022geodiff} &0.3731 \cite{xu2022geodiff} &67.16 \cite{xu2022geodiff} &1.0875 \cite{xu2022geodiff} \\
DGSM** \cite{luo2021predicting}&MPNN, Diffusion &G-QM9, G-Drugs &91.49 &0.2139 &78.73 &1.0154 \\
Torsional** \cite{jing2022torsional}&Diffusion &G-QM9, G-Drugs, G-XL &92.8 &0.178 &72.7* &0.582 \\
\bottomrule
\end{tabular}
\vspace{.1cm}
\scriptsize
\caption{An overview of relevant molecule conformation generation models. All benchmarking metrics are self-reported unless otherwise noted. GEOM-QM9 and GEOM-Drugs are denoted as G-QM9, and G-Drugs for brevity. [*] Torsional Diffusion uses a threshold of 0.75 \r{A} instead of 1.25 \r{A} to calculate coverage on GEOM-Drugs, leading to a deflated score. Notably, Torsional Diffusion outperforms GeoDiff and GeoMol when tested on this threshold. [**] represents the current SOTA.}\label{tab:MolConformation}
\vspace{-.4cm}
\end{table*}

As shown in Table \ref{tab:MolConformation}, Torsional Diffusion appears to outperform all other models, with DGSM and GeoDiff also exhibiting competitive performance. Notably, methods that focus on torsional angles seem to outperform other methods with similar architectures: GeoMol outperforms distance-graph methods like GraphDG and CGCF, while Torsional Diffusion outperforms GeoDiff. 

\subsubsection{Protein Representation Learning}
\paragraph{Overview}
Protein representation learning involves learning embeddings to convert raw protein data into latent space representations, extracting meaningful features and chemical attributes. More specifically, given a protein $x = [ o_1, o_2, \dots o_L ] $, where each $o_1$ represents an amino acid (sequence-based) or atom coordinate (structure-based), learn an embedding $z = [h_1, h_2, \dots h_L]$, where each $h_i \in \mathbb{R}^d$ represents a $d-$dimensional token representation for amino acid $o_i$. This general framework can be extended to a multitude of tasks--- for token property prediction tasks, additionally allocate a prediction $p(y_i | o_i)$ in the embedding, and for pairwise prediction tasks, allocate a prediction $p(y_{ij} | o_i, o_j)$. For overall protein property prediction, allocate predictions $p(y | x)$ or $p(y | x_1, x_2)$ for individual and protein-protein properties, respectively. 

These embeddings can be seen as creating ``richer" data spaces for various models to work on; whereas raw atom graphs and amino acid sequences are entirely arbitrary, these embeddings are aimed to capture various chemical traits and properties (i.e. two \mg{proteins} with similar properties should be represented with similar embeddings, even if their raw forms are very different). Thus, representation learning models do not directly perform any tasks on their own, but instead are incorporated in other generative models to improve their ability to model these important chemical features. 

\paragraph{Datasets} \label{ProteinSequenceRepData} \label{ProteinStructureRepData}
Note that a plethora of datasets exist for all various downstream tasks; we do not describe all such datasets. We focus on datasets used for pre-training in protein representation learning, described below. Note that UniRef, UniParc, and ProteinKG are typically used for sequence-based learning, and PDB and AlphafoldDB are primarily used for structure-based learing. Pfam has been used in both types of models. 
\begin{itemize}
\item \textbf{UniRef} \cite{apweiler2004uniprot} - A clustered version of the Unified Protein KnowledgeBase (UniProtKB), part of the central resource UniProt, which is a curated and labeled set of protein sequences and their functions
\item \textbf{UniParc} \cite{apweiler2004uniprot} - A larger dataset of protein sequences, part of the central resource Uniprot, which includes UniProtKB and adds proteins from a variety of other sources
\item \textbf{ProteinKG} \cite{zhang2022ontoprotein} - A dataset created by Zhang et al. (OntoProtein), aligning biological knowledge with protein sequences with knowledge graphs (KG), used for directly injecting biological knowledge in the representation learning process
\item \textbf{PDB} \cite{berman2000protein} - \textit{Protein Data Bank}, a central archive for all experimentally determined protein structures, widely used in almost all protein structure-related tasks.
\item \textbf{AlphaFoldDB} \cite{varadi2022alphafold} - \textit{AlphaFold Protein Structure Database}, a large dataset created by using AlphaFold2 to predict structures from sequence datasets, such as UniProt and Swiss-Prot.
\item \textbf{Pfam} \cite{mistry2021pfam} - A collection of protein families, used in multiple sequence alignments (MSA)
\end{itemize}

\paragraph{Tasks}
Representation learning models do not directly address tasks on their own but rather supplement a wide range of tasks by refining a latent space. Thus, the protein representation learning field in particular lacks uniformity in testing and training methods, making it hard to make generalized evaluations. Some methods like Tasks Assessing Protein Embeddings (TAPE) \cite{rao2019evaluating} and, more recently, Protein sEquence undERstanding (PEER) \cite{xu2022peer} seek to address this by compiling various testing metrics into a standardized benchmark for evaluating these language models. A few of the most prominent tasks include: 
\begin{itemize}
\item \textbf{Contact Prediction -} Given two amino acid residues, predict the probability that they ``contact", or are within some 3D distance threshold. Measured using precision of top $L/5$ medium-range and long-range contacts (contact between sequentially distant amino acids) according to CASP standards \cite{moult2018critical}.

\item \textbf{Fold Classification -} Also known as remote homology prediction. Given an amino acid sequence, predict the corresponding fold class \cite{schaeffer2011protein}. Used for finding structural similarities for distantly related inputs. Tested using the dataset proposed by Hou et al. \cite{hou2018deepsf}, which comprises three versions of increasing difficulty: Family (test set in the same family as the training set), Superfamily (same superfamily as training) and Fold (same fold as training). Measured using accuracy. \label{FoldClass}

\item \textbf{Stability Prediction -} Given a protein, output a label $y \in \mathbb{R}$ representing the most extreme circumstances under which a protein maintains its fold above a concentration threshold. Measured using Spearman's $\rho$.

\item \textbf{PPI -} \textit{Protein-Protein Interaction}, given two proteins, predict whether or not they interact (binary classification). Measured using accuracy. 
\end{itemize}

\paragraph{Metrics}
The following metrics are used to evaluate the above tasks:
\begin{itemize}
    \item \textbf{Accuracy -} Ratio between correct classifications to total classifications.
    \item \textbf{Precision -} Accuracy within the set of selected elements; in contact prediction, this refers to the following: of the top $L/5$ contacts with the highest predicted probability, find the percentage of these which are ground-truth contacts. $L$ refers to the length of a protein sequence.
    \item \textbf{Spearman's $\mathbf{\rho}$ - } \textit{Spearman's rank correlation coefficient}, found by sorting the ground-truth and predicted output values into ranked order (so each protein has a true rank $y_i \in \mathbb{N}$ and predicted rank $x_i \in \mathbb{N}$) and computing: $${\displaystyle \rho=1-{\frac {6\sum d_{i}^{2}}{n(n^{2}-1)}},}$$
    where $d_i$ refers to the numerical difference in ranks for each $y_i, x_i$, and $n$ refers to the total number of elements in each set.
\end{itemize}

\paragraph{Models}

\begin{table*}[!htp]\centering
\scriptsize
\begin{tabular}{lccccccc}\toprule
\multirow{2}{*}{\textbf{Model}} &\multirow{2}{*}{\textbf{Type of Model}} &\multirow{2}{*}{\textbf{Dataset}} &\multirow{2}{*}{\textbf{Contact ($\uparrow$)}} &\multicolumn{3}{c}{\textbf{Fold Classification ($\uparrow$)}} & \multirow{2}{*}{\textbf{Stability} ($\uparrow$)} \\ & & & & \textbf{Family} & \textbf{Superfam.} & \textbf{Fold} & \\\midrule
UniRep \cite{alley2019unified} &LSTM RNN &UniRef50 &\textbackslash &\textbackslash &\textbackslash &\textbackslash &\textbackslash \\
ProtBERT \cite{elnaggar2021prottrans}&BERT &BFD100, UniRef100 &0.556 \cite{ma2023retrieved} &0.528 \cite{ma2023retrieved} & 0.192 \cite{ma2023retrieved}& 0.170 \cite{ma2023retrieved}& 0.651 \cite{ma2023retrieved} \\
ESM-1B \cite{rives2021biological}&Transformer &UniParc &0.458 &0.978 & 0.601 & 0.268 \cite{zhang2022protein}&0.695 \cite{zhang2022protein}\\
MSA Trans. \cite{rao2021msa}&Transformer &26M MSAs &0.618 \cite{ma2023retrieved} &0.958 \cite{ma2023retrieved} & 0.503 \cite{ma2023retrieved} & 0.235 \cite{ma2023retrieved}&0.796 \cite{ma2023retrieved} \\
RSA** \cite{ma2023retrieved}&Transformer &Pfam &0.717 &0.987 & 0.677 & 0.267 & 0.987 \\
OntoProtein** \cite{zhang2022ontoprotein}&ProtBERT, BERT/Trans.&ProteinKG* &0.40  &0.96 \cite{ma2023retrieved} & \textbackslash & 0.24 &0.75 \cite{ma2023retrieved} \\
KeAP** \cite{zhou2023protein}&BERT, Transformer &ProteinKG &0.62 &\textbackslash & \textbackslash & 0.29 &0.82 \\
GearNET** \cite{zhang2022protein}&Geo-EGNN &AlphaFoldDB & \textbackslash & 0.995 & 0.703 & 0.483 & \textbackslash
\\
DeepFRI \cite{gligorijevic2021structure}&LSTM, GCNN &Pfam &\textbackslash &0.732 \cite{hermosilla2020intrinsic} & 0.206 \cite{hermosilla2020intrinsic} & 0.153 \cite{hermosilla2020intrinsic} & \textbackslash
\\ 
IEConv \cite{hermosilla2020intrinsic}&CNN &PDB & \textbackslash & 0.997 & 0.806 & 0.503 & \textbackslash
\\
\bottomrule
\end{tabular}
\vspace{0.05cm}

\caption{An overview of the most relevant sequence-based protein representation learning models. All metrics are self-reported unless otherwise noted; metrics reported by different papers may be incomparable due to different test settings. Note that OntoProtein and KeAP have two separate architectures for protein, and knowledge encoding respectively. [**] denotes the current SOTA.}\label{tab:SeqProtRepLearning}

\end{table*}

As mentioned above, protein representation learning models can be divided into two broad categories: sequence-based and structure-based. This distinction simply refers to the type of input these models learn representations for; we begin by discussing sequence-based learning models. Both amino acid sequences and protein structures carry valuable information about a protein's function and binding properties, and thus these two types of learning can be seen as complements to each other.   

Unified Representation (UniRep) \cite{alley2019unified} used a multiplicative long-short-term memory recurrent neural network (mLSTM RNN) to approach the task of representation learning. By iterating through each amino acid and comparing its prediction with the true amino acid type, adjusting its parameters at each step, UniRep gradually learns a better representation of the sequence. While the LSTM model has been proven inferior to newer transformer and natural language models, the basic concepts laid an important groundwork for future representation learning models.

ProtBERT \cite{elnaggar2021prottrans} improved upon performance by applying a BERT model to amino acid sequences, treating each residue as a word. BERT stands for Bidirectional Encoder Representations from Transformers, which is a pre-trained language model for neural language processing. Several variations, such as RoBERTa and DeBERTa, improve upon BERT by enhancing parameters and attention mechanisms and are used in models discussed later. ESM-1B \cite{rives2021biological} trained a deep transformer with 33 layers and 650 million parameters, using the masked language modeling (MLM) objective for training. ESM-1b considered residue contexts across the entire sequence through its stacked attention layers, and its complexity and size allowed it to outperform other SOTA models on a wide range of tasks, as shown below. ESM-1b's massive size also resulted in expensive training costs, however, and smaller research groups would had to find more unique approaches to compete with ESM-1b on representation learning. 

Following ESM-1b's breakthrough, a series of models applied the concept of direct knowledge injection. Previous models had simply treated these sequences as generic sequences of tokens, failing to take advantage of the vast biological knowledge in existence today. One avenue for knowledge injection comes from the analysis of Multiple Sequence Alignments (MSA), which groups proteins concerning their evolutionary families. Because inherent patterns appear as a result of the evolutionary process, analyzing MSAs can result in a richer understanding of protein sequences. MSA Transformer \cite{rao2021msa} was the first model to intersect a protein language model with MSA analysis, pretraining a transformer model on MSA input. While MSAs can provide large benefits, the alignment process is computationally costly. Retrieved Sequence Augmentation (RSA) \cite{ma2023retrieved} provides a solution to this, using a dense sequence retriever to directly augment input proteins with corresponding evolutionary information from the Pfam database. This allows RSA to incorporate similar information as traditional MSA methods, without the computationally costly MSA alignment process. OntoProtein \cite{zhang2022ontoprotein} incorporates direct knowledge injection by pretraining on gene ontology (GO) knowledge graphs (KG). OntoProtein uses the ProtBERT architecture for encoding protein sequences but introduces a second encoder for training on the GO input. To perform this hybrid encoding, Zhang et al. \cite{zhang2022ontoprotein} create a new dataset, ProteinKG25, which aligns protein sequences with their respective GO annotations in text format. Knowledge-exploited Auto-encoder for Protein (KeAP) \cite{zhou2023protein} extends upon this by exploring protein and annotated text on a more granular level; while OntoProtein models relationships over proteins and texts as a whole, KeAP uses a cross-attention mechanism to perform token-level exploration on individual amino acids and words. 

Now, we discuss a few structure-based learning models. IEConv \cite{hermosilla2020intrinsic} incorporates both intrinsic distance (based on connectivity in a node-graph representation) and extrinsic distance (based on physical 3D distance) information in a convolutional neural network (CNN), using contrastive learning to pre-train. Additionally, IEConv uses protein-specific pooling methods, like reducing each amino acid to its alpha carbon, to reduce computational complexity and allow for more learned features per atom under memory constraints. DeepFRI \cite{gligorijevic2021structure} combines the idea of sequence representation learning with structural representation learning, incorporating both a pre-trained LSTM to learn sequence data and a GCN to learn from contact map input (matrix representing 3D distances between residues). GearNET \cite{zhang2022protein} applies a GNN to a graph of residues, with three types of directed edges defined for sequential/physical distance and K-nearest neighbors. While previous models had only considered message-passing between residues, GearNET applied direct message-passing between edges in its variant model GearNET-Edge, which led to even further improved performance on various metrics. In addition, GearNET-IEConv added an intrinsic-extrinsic convolutional layer inspired by Hermosilla et al. \cite{hermosilla2020intrinsic} to improve performance, with GearNET-Edge-IEConv combining both of these features.

\subsection{Antibody} 
\subsubsection{Antibody Task Background} \label{antibodytaskbackground}

Antibodies are Y-shaped proteins used by the immune system to identify and neutralize foreign objects like bacteria and viruses. Each antibody contains two identical sets of chains, with each set containing one heavy chain and one light chain. The variable domains of these chains control binding specificity for various antigens, containing the binding surface, known as the paratope. The paratope binds with specificity to a target antigen at its respective binding region, known as the epitope. The paratope is typically located within six Complementarity-Determining Regions (CDRs) on the antibody: three from the light chain (CDR-L1, CDR-L2, CDR-L3) and three from the heavy chain (CDR-H1, CDR-H2, CDR-H3). The Complementary-Determining Region 3 (CDR-H3) is the most diverse in terms of amino acid sequence and length--- due to its lack of constraints, it is often the most complex to produce, making its generation the primary focus for modern generative models. 
The end goal of machine learning is to generate antibodies in silico. In approaching this goal, the typical pipeline often includes an amino acid sequence input, sequence representation learning, structure prediction, paratope-epitope prediction, antibody-antigen docking, CDR generation, and evaluation, typically in the form of affinity prediction. The final output is a generated antibody, either in the form of a 1D sequence, 3D structure, or both. Sometimes, a 3D structure is input directly, with the sequence being predicted from the structure. Machine learning methods have been applied to various fragments of the process, with some more recent models \cite{kong2022conditional}, 
\cite{kong2023end}, \cite{jin2021iterative} handling both sequence and structure generation simultaneously, and others \cite{kong2023end} even addressing the entire end-to-end process. Other subtasks like drug-applicable attribute prediction and affinity prediction have been explored using predictive AI models, but these tasks are not generative by nature and not be discussed in detail. 

\subsubsection{Antibody Representation Learning} \label{AntibodyRepLearning}
\paragraph{Overview}
This task is identical to the protein sequence representation learning, but language models in this field pretrain on antibody-specific data, in particular, outperforming general protein representation models on antibody-based tasks. 

\begin{table*}[!htp]\centering
\scriptsize
\begin{tabular}{lcccl}\toprule
\textbf{Model} &\textbf{Type of Model} &\textbf{Dataset}  &\textbf{AAR (\%, $\uparrow$)} &\textbf{Additional Self-Benchmarking Notes} \\\midrule
BERTTransformer \cite{li2022antibody}&BERT &OAS  &\textbackslash & Improved binding affinity prediction over generic CNN \\ 
AntiBERTy \cite{ruffolo2021deciphering}&BERT &OAS &26.00 \cite{gao2023pre} &Reveals evolutionary trajectories for affinity maturation\\ 
AbLang \cite{olsen2022ablang} &RoBERTa &OAS &33.60 \cite{gao2023pre} &Clustering better represents distinction between B-cell type\\
& & & & and other metrics (compared to ESM-1b) \\ 
PARA \cite{gao2023pre}&DeBERTa &OAS&34.20 \cite{gao2023pre} &Improved heavy/light chain matching (compared to AntiBERTy) \\ 
AntiBERTa \cite{leem2022deciphering}&RoBERTa &OAS & \textbackslash &Clustering better represents distinction between B-cell type \\ 
& & & & (compared to ProtBERT), improves paratope prediction \\
\bottomrule
\end{tabular}
\vspace{0.05cm}

\caption{An overview of relevant antibody representation learning models. Due to the lack of standardization in benchmarking metrics, we note each model's benchmarking findings.}\label{tab:AntibodyRepLearning}

\end{table*}

\paragraph{Datasets}
All reported methods pre-train on the OAS dataset. In all methods, OAS was divided between heavy and light chains, with some \cite{gao2023pre, olsen2022ablang} also using Linclust to cluster the data. Leem \textit{et. al} also includes pre-training on the general Pfam database for protein sequences, as a comparison to pre-training on the antibody-specific OAS dataset. 
\begin{itemize}
    \item \textbf{OAS} \cite{olsen2022observed} - \textit{Observed Antibody Space}, a compilation of over one billion raw antibody sequences.
\end{itemize}
\paragraph{Tasks}
Given the variety of applications for representation learning, the benchmarks used for this task vary widely; some use Pearson correlation to measure affinity binding prediction \cite{leem2022deciphering}, while others \cite{gao2023pre} use amino acid recovery (AAR) to measure accuracy in a CDR-H3 sequence prediction task. Some use a masked language modeling task (MLM) to measure the quality of BCR sequence representation \cite{leem2022deciphering}, while others measure accuracy in paratope prediction \cite{leem2022deciphering, ruffolo2021deciphering}. However, we do not have a standardized benchmarking system as we did with protein representation learning, so we only report CDR-H3 sequence prediction in Table \ref{tab:AntibodyRepLearning}, which will be elaborated on more in the Antibody CDR Generation section (see page \pageref{AntibodyCDR}). Other evaluations are left as individual notes for each model. 
\paragraph{Metrics}
As noted above, we only report results on the CDR-H3 prediction task, which is measured with amino acid recovery (AAR), which represents the percentage of matching amino acids between ground truth and generated amino acid sequences.

\paragraph{Models}

Li et al. \cite{li2022antibody} created the BERTTransformer model, serving as the first representation learning model aimed specifically at antibody learning. Li et al. find that pre-training on Pfam results in improved performance on affinity binding prediction when compared to pre-training on OAS, suggesting some merit to general protein datasets in pre-training for antibody representation learning. 
AntiBERTy \cite{ruffolo2021deciphering} is a BERT-based model, and produces valuable embeddings used in structure prediction by IgFold \cite{ruffolo2023fast}. AntiBERTa \cite{leem2022deciphering} and AbLang \cite{olsen2022ablang} both use the more optimized RoBERTa architecture, while the most recent model PARA \cite{gao2023pre} uses the current state-of-the-art DeBERTa architecture, and outperforms previous pre-trained models in AAR.

A comprehensive overview of representation learning models can be seen in Table \ref{tab:AntibodyRepLearning}. As representation learning serves as a general tool, rather than the optimization for a specific task, these methods all use a variety of self-chosen applications to measure performance. PARA tests amino acid recovery against AbLang and ` on the CDR-H3 prediction task, reporting improved accuracy. AntiBERTy and AntiBERTa report improved paratope prediction, while AbLang and AntiBERTa focus on improved clustering to distinguish between naive and memory B-Cells. AntiBERTa also explores clustering quality in terms of drug-applicable traits, such as origin, closest human V gene identity, and anti-drug antibody (ADA) response scores.

\subsubsection{Antibody Structure Prediction} \label{AntibodyStructPredict}
\paragraph{Overview}
Antibody structure prediction extends general protein structure prediction methods, with the key difference lying in the reliance on multiple sequence alignment (MSA) of homologous proteins mapping evolutionary relationships between genetically related sequences. \mggg{Because the relevant evolutionary histories for CDR-H3 loop sequences are lacking, MSAs in antibodies are not always available.} This makes models like AlphaFold \cite{jumper2021highly}, designed for general protein structure prediction, highly inefficient and slow. Thus, antibody-specific structural prediction methods seek to predict antibody structure without the need for an input MSA \cite{ruffolo2023fast}. 
\paragraph{Datasets} \label{AntibodyStructurePredict}
Models in this field all train on the Structural Antibody Database (SAbDab), with the Rosetta Antibody Benchmark (RAB), used as a validation set in some cases \cite{abanades2022ablooper, abanades2023immunebuilder}. Note that ``canonical conformations" do not exist for the highly variable H3 region due to high variability, which remains a challenge to structurally model. 
\begin{itemize} 
    \item \textbf{SAbDab \cite{dunbar2014sabdab} -} \textit{Structural Antibody Database}, a collection of all antibody structures in PDB \cite{berman2000protein}. Structures are annotated with antibody-specific structural data like canonical conformations for complementarity determining regions (CDR), orientation between the variable domains on the light and heavy chains, and the presence of constant domains in the structure. 
    \item \textbf{RAB \cite{adolf2018rosettaantibodydesign} -} \textit{Rosetta Antibody Benchmark}, a hand-selected set of 60 antibody-antigen complexes, chosen to be as diverse in CDR lengths and clusters as possible. 
\end{itemize}%TODO: discuss redundancy filtering

\paragraph{Task}
Given an antibody amino acid sequence (organized into heavy chain, light chain, and linker sequences), generate a set of 3D point coordinates for each amino acid residue. 
\paragraph{Metrics}
For benchmarking, models compare similarities between the predicted and ground truth structures through the following  metrics:
\begin{itemize}
    \item \textbf{RMSD} - \textit{Root-mean-square deviation}, measures distances between ground truth and generated residue coordinates. Also used in protein structure prediction (see page \pageref{ProtStructurePrediction})
    \item \textbf{OCD} \cite{marze2016improved} - \textit{Orientational coordinate distance}, measures similarity between light-heavy orientational coordinates (LHOC) of ground truth and generated residues. LHOC is defined by Marze et al. \cite{marze2016improved}, consisting of four metrics to measure orientation between light and heavy chains.
\end{itemize}
used root-mean-square deviation (RMSD) to measure the discrepancy between predicted and ground truth 3D structures, while some models \cite{ruffolo2023fast}, \cite{ruffolo2022antibody}, \cite{wu2022tfold} also used orientational coordinate distance (OCD) to measure the accuracy of predicted relative position between heavy and light chains in some cases. 
% TODO: 
% Hit Enter to reply
% eliannak: do you mean between heavy and light chain? i thought the OCD was measured between the predicted and actual - not sure measuring the distance between heavy and light chains make sense

% Jul 5, 2023 8:50 PM
% You: predicted relative position between heavy/light chains vs ground truth relative position between heavy/light chains

% Sep 6, 2023 8:14 PM • Edit • Delete
% You: will reformat into bullet points
\paragraph{Models}
\begin{table*}[!htp]
\centering

\scriptsize
\begin{tabular}{lcccccc}\toprule
\textbf{Model} &\textbf{Type of Model} &\textbf{Dataset} &\textbf{H3 RMSD} (\r{A}, $\downarrow$)  &\textbf{OCD ($\downarrow$)}  &\textbf{Generation Time} (s, $\downarrow)$)  \\\midrule
IgFold** \cite{ruffolo2023fast} &BERT, Graph &OAS, SAbDab &2.99 &3.77 \cite{ruffolo2023fast}&0.46 \cite{ruffolo2023fast} & \\ 
DeepAB \cite{ruffolo2022antibody} &LSTM, NN &OAS, SAbDab &3.28 &3.6 \cite{ruffolo2023fast}&1620.22 \cite{wu2022tfold}, ~600 \cite{ruffolo2023fast} \\ 
ABLooper \cite{abanades2022ablooper} &EGNN &SAbDab, RAB&3.2 &4.53 \cite{ruffolo2023fast} &~30-60 \cite{ruffolo2023fast}  \\ 
ImmuneBuilder \cite{abanades2023immunebuilder} &AlphaFold - Multimer &SAbDab, RAB&2.81 &4.9 \cite{ruffolo2023fast}&~5*   \\ 
tFold-Ab** \cite{wu2022tfold} &AlphaFold - Multimer &SAbDab&2.74 &3.21 \cite{ruffolo2023fast}&2.23* \cite{wu2022tfold}  \\
xTrimoABFold \cite{wang2022xtrimoabfold} &AlphaFold - Multimer  &PDB&1.25  & \textbackslash & \textbackslash\\
\bottomrule
\end{tabular}
\vspace{.1cm}
\caption{An overview of the most relevant antibody structure prediction models. All benchmarking metrics are self-reported unless otherwise noted. Igfold and tFold-Ab use slightly different benchmarking methods, so their results may not be comparable. [*] denotes significant discrepancies which result from varying benchmarking methods. [**] denotes the current SOTA.} 
\label{tab:AntibodyStructPredict}
\end{table*}

tFold-Ab \cite{wu2022tfold}, xTrimoABFold \cite{wang2022xtrimoabfold}, and ABodyBuilder2 \cite{abanades2023immunebuilder} all apply similar methods to AlphaFold, replacing the unnecessary MSA searching component. AbLooper \cite{abanades2022ablooper} uses five EGNNS to each individually predict CDR structures, outputting the positional average of the five structures--- this allows for the evaluation of confidence as the deviation between these predicted structures. Some models do not predict structures directly: DeepH3 \cite{ruffolo2020geometric} uses a deep residual neural network to first predict structural restraints, using Rosetta to then output atomic structures. SimpleDH3 \cite{zenkova2021simple} adds ELMo embeddings to DeepH3, and DeepAb \cite{ruffolo2022antibody}, the most recent version, adds interpretable attention layers to enhance output from the neural network. 

Igfold \cite{ruffolo2023fast} takes advantage of AntiBERTy's \cite{ruffolo2021deciphering} sequence embeddings to predict antibody structures using invariant point attention. While some other models, such as tFold-Ab \cite{wu2022tfold}, xTrimoABFold \cite{wang2022xtrimoabfold}, and ABodyBuilder2 \cite{abanades2023immunebuilder}, demonstrate more accurate structures, IgFold produces structures at a superior speed, making it the established state-of-the-art model. Igfold's downside is its reliance on PyRosetta to produce side-chain conformations, as the model Igfold itself only outputs backbone atoms. This adds significant time to the full process, as it takes around 0.46 seconds for Igfold to produce backbone atoms, but an additional 21.86 seconds, on average, for PyRosetta to convert these into full-atom structures \cite{wu2022tfold}. By contrast, tFold-Ab \cite{wu2022tfold} generates full-atom structures in less time than the Igfold + PyRosetta pipeline (2.23 seconds compared to 22.32 seconds). 

Notably, in the training process, ImmuneBuilder and xTrimoABFold choose to include antibodies with identical sequences to expose models to antibodies with multiple conformations (identical sequences with different structures). On the other hand, DeepAB includes an entirely non-redundant set of sequences, seeking to expose its model only to unique items.  

Table \ref{tab:AntibodyStructPredict} presents detailed results for the performance of the discussed models. While IgFold does not have the absolute highest performance in terms of RMSD and OCD when compared to other methods, its speed makes it the state-of-the-art model. This speed advantage is contested only by tFold-Ab, which generates full-atom structures in less time than IgFold, as it circumvents IgFold’s reliance on computationally expensive Rosetta energy functions for side-chain prediction. Note that discrepancies in time results may result from varying benchmarking methods--- time results from tFold-ab were reported on their own SAbDab-22H1-Ab/Nano dataset, while time results from IgFold were reported using their IgFold-Ab benchmark. 

%Please add the following packages if necessary:
%\usepackage{booktabs, multirow} % for borders and merged ranges
%\usepackage{soul}% for underlines
%\usepackage[table]{xcolor} % for cell colors
%\usepackage{changepage,threeparttable} % for wide tables
%If the table is too wide, replace \begin{table}[!htp]...\end{table} with
%\begin{adjustwidth}{-2.5 cm}{-2.5 cm}\centering\begin{threeparttable}[!htb]...\end{threeparttable}\end{adjustwidth}

\subsubsection{Antibody CDR Generation} \label{AntibodyCDR}

\begin{figure*}
    \centering
    \includegraphics[width = \textwidth ]{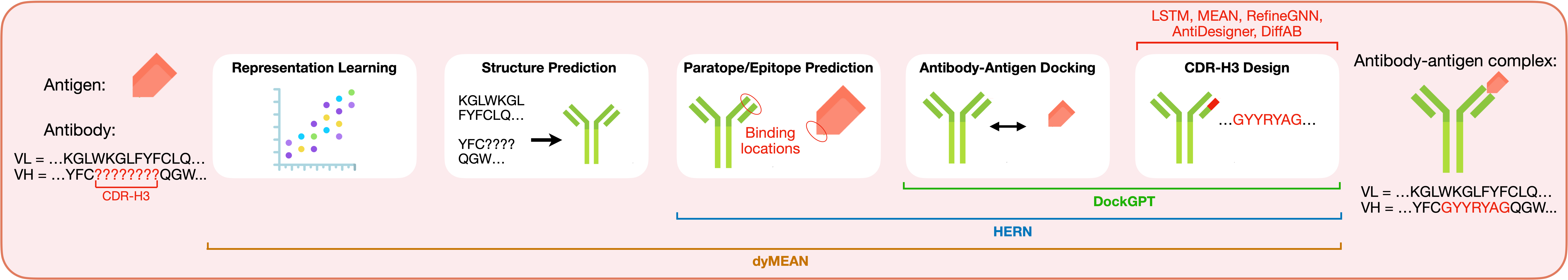}
    \caption{A comprehensive overview of the antibody generation pipeline for CDR-H3 design \cite{Silva_2021, zhao2018silico, bocharov2018basic, stanzionechapter, ambrosetti2020protocol}. The inputs are a target antigen and antibody information (without CDR-H3), and the output is an antibody-antigen complex with a designed CDR-H3 sequence. Note that while most antibody CDR-H3 generation methods only generate the CDR-H3 region, needing a docked structure as input, some methods like DockGPT \cite{mcpartlon2023deep}, HERN \cite{jin2022antibody}, and dyMEAN \cite{kong2023end} perform multiple steps of the pipeline on their own. 
    %Images taken from \cite{Silva_2021, zhao2018silico, bocharov2018basic, stanzionechapter, ambrosetti2020protocol}.
    }
\end{figure*}

\paragraph{Overview}
CDR generation involves the generation of the variable regions of the antibody, directly determining binding affinity and making it the core of antibody generation. As mentioned above, the CDR-H3 region is the most variable, making it the most difficult region for models to generate and thus the primary region of focus. 
\paragraph{Datasets}
The primary datasets used in recent models are listed below. Note that SAbDab is the main dataset typically used for training, whereas RAbD and SKEMPI are specialized datasets used for testing specific downstream tasks.
\begin{itemize}
\item \textbf{SAbDab \cite{dunbar2014sabdab} -}  \textit{The Structural Antibody Database}, an annotated dataset of all antibody structures contained in PDB \cite{berman2000protein}, used in structure prediction (see page \pageref{AntibodyStructurePredict}) 
\item \textbf{RAB \cite{adolf2018rosettaantibodydesign} -} \textit{Rosetta Antibody Benchmark}, a hand-selected group of 60 diverse antibody-antigen complexes, also used in antibody structure prediction (see page \pageref{AntibodyStructurePredict}). 
\item \textbf{SKEMPI \cite{jankauskaite2019skempi} -} \textit{Structural Database of Kinetics and Energetics of Mutant Protein Interactions}, a dataset of energy changes for mutations of protein-protein interactions.
\end{itemize}
\paragraph{Tasks}
Three general tasks are used to benchmark models in this field:
\begin{itemize}

\item \textbf{Sequence and Structure Modeling -} Given an antibody structure, predict its corresponding sequence (or vice-versa). AAR is used to evaluate generated sequences, while RMSD is used to evaluate generated structures. Testing for this task is done on the SAbDab dataset.
% TODO: fact-check this task? 

\item \textbf{CDR-H3 Generation -} Given a target antigen, generate the binding antibody CDR-H3 region. AAR is used to evaluate sequences, while RMSD and TM-Score are used to evaluate structures. Testing is done on the RAbD dataset.

\item \textbf{Affinity Optimization -} Given an antibody-antigen pair, make changes to the CDR region to optimize its binding affinity with the target antigen. Performance is measured by $\Delta\Delta G$, and testing is done on SKEMPI v2.0.

\end{itemize} 
\paragraph{Metrics}
The following metrics are used to evaluate CDR generation models:
\begin{itemize}
    \item \textbf{AAR -} \textit{Amino acid recovery}, a comparison between ground truth and generated amino acid sequences, also measured in antibody representation learning (see \pageref{AntibodyRepLearning}).
    \item \textbf{RMSD} - \textit{Root-mean-square deviation}, measures distances between ground truth and generated residue coordinates. Also used in protein structure prediction (see page \pageref{ProtStructurePrediction})
    \item \textbf{TM-score} \cite{zhang2004scoring} - \textit{Template Modeling Score}, a distance measurement capturing both local and global similarities, also used in protein structure prediction (see page \pageref{ProtStructurePrediction}).
    \item $\mathbf{\Delta\Delta G}$ - Change in binding energy after affinity optimization. Binding energy is predicted by the geometric network proposed by Shan et al. \cite{shan2022deep}. 
\end{itemize}

\paragraph{Models}
While some approaches to the CDR generation problem have been sequence-based, such as the LSTM used by Akbar et al. \cite{akbar2022silico}, the field has seen movement towards structure-based design and even more recently, sequence-structure co-design, as this allows models to incorporate both 1D amino acid sequence information and 3D structure information in the generation process. RefineGNN \cite{jin2021iterative} was the first to introduce structure/sequence co-design, representing both structural and sequential information in a graph and performing iterative refinement. A Multi-channel Equivariant Attention Network (MEAN) \cite{kong2022conditional} improves efficiency by predicting all amino acids within a CDR region at once, allowing for fewer iterations than RefineGNN's prediction for each amino acid residue. AntiDesigner \cite{cross} implements a Protein Complex Invariant Embedding (PCIE) and dual Multi-Layer Perceptrons (MLPs) to design structures and sequences in a one-shot manner, removing the need for iterative refinement entirely. DiffAb \cite{diffab} defines diffusion processes for both amino acid sequences and 3D backbone coordinates, serving as one of the first DDPMs in the field of antibody generation. 
% HTP~\citep{wu2023hierarchical} 
%
% 
\begin{table*}[!tb]
\centering
\scriptsize
\begin{tabular}{lccccccccc}\toprule
\multirow{2}{*}{\textbf{Model}} &\multirow{2}{*}{\textbf{Type of Model}} &\multirow{2}{*}{\textbf{Dataset}} & \multicolumn{2}{c}{\textbf{Modeling}} & \multicolumn{3}{c}{\textbf{CDR-H3 Design}} & \textbf{Affinity} \\ & & & \textbf{AAR} (\%, $\uparrow$) &\textbf{RMSD} ($\downarrow$) &\textbf{AAR} (\%, $\uparrow$) &\textbf{TM-Score} ($\uparrow$) &\textbf{RMSD} ($\downarrow$)&$\mathbf{\Delta \Delta G} (\downarrow)$\\\midrule
LSTM \cite{akbar2022silico} &LSTM &Self-made &15.69  \cite{kong2022conditional} & \textbackslash &22.36 \cite{kong2022conditional} &\textbackslash &\textbackslash &-1.48  \cite{kong2022conditional} \\
RefineGNN \cite{jin2021iterative} &GNN &SAbDab &21.13 \cite{kong2022conditional} &6 \cite{kong2022conditional} &29.79 \cite{kong2022conditional} &0.8308 \cite{kong2022conditional} &7.55 \cite{kong2022conditional} &-3.98 \cite{kong2022conditional} \\
MEAN \cite{kong2022conditional} &EGNN &SAbDab &36.38 &2.21 &36.77&0.9812 &1.81 &-5.33 \\
AntiDesigner** \cite{cross} &MLP &SAbDab &37.37 &1.97 &40.94 &0.985 &1.55 &-10.78 \\
DiffAB \cite{diffab} &Diffusion &SAbDab &26.78 &3.597 & 35.31 \cite{kong2023end} &0.9695 \cite{kong2023end} &\textbackslash &-2.17 \cite{kong2023end} \\
DockGPT \cite{mcpartlon2023deep} & Transformer & SAbDab & \textbackslash & \textbackslash & \textbackslash & \textbackslash & 1.88 & \textbackslash \\
HERN \cite{jin2022antibody} & EGNN & SAbDab & \textbackslash & \textbackslash & 34.1 & \textbackslash & \textbackslash & \textbackslash \\
dyMEAN** \cite{kong2023end} &EGNN &SAbDab &\textbackslash & \textbackslash&43.65&0.9726 &\textbackslash &-7.31 \\
% HTP~\citep{wu2023hierarchical} & E(3)-eGNN &SAbDab \\ 
\bottomrule
\end{tabular}
\vspace{0.05cm}
\caption{A summary of relevant CDR generation models. All benchmarking metrics are self-reported unless otherwise noted. dyMEAN and DiffAB use slightly different evaluation conditions, so their results may not be fully comparable. [**] denotes the current SOTA.}\label{tab: AntibodyCDR}
\end{table*} 

While the above models all show significant progress in the specific CDR generation task, these models still require the input of fully docked antibody-antigen structures; thus, in real-world applications where only sequence and/or structure data is available, these models must be paired with structure prediction and docking methods to operate. As a result, some methods simultaneously implement other parts of the antibody generation process in their model. The deep learning method DockGPT \cite{mcpartlon2023deep} includes an adaptation of their protein-protein docking method for antibody CDR-H3 docking and design. By adding an encoding feature to distinguish CDR residues during training and providing structures with missing CDR regions as input, DockGPT simultaneously docks and designs all CDR regions. HERN \cite{jin2022antibody} uses hierarchical equivariant refinement to both generate and dock paratopes for target antigens. However, HERN requires a defined epitope region as input and needs to be combined with epitope prediction models to dock with antigens whose epitopes are not identified. Dynamic Multichannel Equivariant Graph Network (dyMEAN) \cite{kong2023end} extends these ideas even further by creating the first end-to-end method, incorporating structure prediction, docking, and CDR generation in a singular model. To demonstrate its improved performance, dyMEAN benchmarks against pipelines combining models for each subtask (Igfold $\Rightarrow$ HDock $\Rightarrow$ MEAN to perform structure prediction, docking, and CDR generation, for example). 

%Please add the following packages if necessary:
%\usepackage{booktabs, multirow} % for borders and merged ranges
%\usepackage{soul}% for underlines
%\usepackage[table]{xcolor} % for cell colors
%\usepackage{changepage,threeparttable} % for wide tables
%If the table is too wide, replace \begin{table}[!htp]...\end{table} with
%\begin{adjustwidth}{-2.5 cm}{-2.5 cm}\centering\begin{threeparttable}[!htb]...\end{threeparttable}\end{adjustwidth}

\subsubsection{Peptide}
\paragraph{Signal Peptide Prediction}
\bib{Signal peptides play a crucial role in protein translocation, but predicting these regions remains difficult due to the lack of clearly defined motifs. To address this issue, PEFT-SP \cite{zeng2023peft} takes advantage of the abilities of existing large protein language models like ESM2. PEFT-SP uses efficient fine-tuning methods like Prompt Tuning \cite{lester2021power} and Low-Rank Adaptation (LoRA) \cite{hu2021lora} to fine-tune ESM2 on an annotated signal peptide dataset created for SignalP 6.0 \cite{teufel2022signalp}, the previous SOTA model. For benchmarking, performance was measured separately for each of the four organism groups in the dataset: Archaea, Eukarya, Gram-positive, and Gram-negative bacteria. PEFT-SP outperforms SignalP 6.0 and other models over the majority of signal peptide prediction categories.}

\end{document}

%% file: fig_tree.tex
\definecolor{mycolor}{RGB}{135, 200, 215}
\definecolor{mycolor2}{RGB}{215, 145, 225}

\tikzstyle{my-box}=[
    rectangle,
    draw=hidden-draw,
    rounded corners,
    text opacity=1,
    minimum height=1.5em,
    minimum width=5em,
    inner sep=2pt,
    align=center,
    fill opacity=.5,
    line width=0.8pt,
]
\tikzset{
leaf/.style={
my-box,
minimum height=1.5em,
fill=mycolor, % Change the RGB values to the desired values
text=black,
align=left,
font=\footnotesize,
inner xsep=2pt,
inner ysep=4pt,
line width=0.8pt
}
}
\begin{figure*}[t!]
    \centering\scriptsize
    \begin{adjustbox}{width=0.7\textwidth}
        \begin{forest}
            forked edges,
            for tree={
                grow=east,
                reversed=true,
                anchor=base west,
                parent anchor=east,
                child anchor=west,
                base=center,
                font=\large,
                rectangle,
                draw=hidden-draw,
                rounded corners,
                align=left,
                text centered,
                minimum width=6em,
                edge+={darkgray, line width=1pt},
                s sep=3pt,
                inner xsep=2pt,
                inner ysep=3pt,
                line width=0.8pt,
                ver/.style={rotate=90, child anchor=north, parent anchor=south, anchor=center},
            },
            where level=1{text width=6em,font=\normalsize,}{},
            where level=2{text width=9em,font=\normalsize,}{},
            where level=3{text width=9em,font=\normalsize,}{},
            where level=4{text width=7em,font=\normalsize,}{},
            [
                Generative Drug Design
                [
                   Molecule
                   [
                   Target-Agnostic \\ Generation
                        [
                            Datasets
                            [
                               QM9 (Ramakrishnan et al.{,} 2014){,} \\ GEOM-Drugs (Axelrod et al.{,} 2022),leaf, text width=33em
                            ]
                        ]
                        [
                            Metrics 
                            [
                                Atom Stability{,} Molecule Stability{,} Validity{,} \\ Uniqueness{,} Novelty{,} QED (Bickerton et al.{,} 2012), leaf, text width=33em
                            ]
                        ]
                        [
                            Models 
                            [
                                CVAE (Gomez et al.{,} 2018){,} GVAE (Kusner et al.{,} 2017){,} \\ SD-VAE (Dai et al.{,} 2018){,} JTVAE (Jin et al.{,} 2018){,}  \\ E-NF (Satorras et al.{,} 2021){,} G-SchNet (Gebauer et al.{,} 2019){,}  \\ EDM (Hoogeboom et al.{,} 2022){,} GCDM (Morehead et al.{,} 2023){,} \\ MDM (Huang et al.{,} 2022){,} GeoLDM (Xu et al.{,} 2023){,} \\ JODO (Huang et al.{,} 2023){,} MiDi (Vignac et al.{,} 2023), leaf, text width=33em
                            ]
                        ]
                    ]
                    [
                        Target-Aware  \\ Generation
                        [
                            Datasets 
                            [
                                CrossDocked2020 (Francoeur et al.{,} 2020){,} \\ ZINC20 (Irwin et al.{,} 2020){,} Binding MOAD (Hu et al.{,} 2005), leaf, text width=33em
                            ]
                        ]
                        [
                            Metrics 
                            [
                                AutoDock Vina (Trott et al.{,} 2010){,} High Affinity Percentage{,} \\ QED (Bickerton et al.{,} 2012){,} SAScore (Ertl et al.{,} 2009){,} 
                                \\ Diversity, leaf, text width=33em
                            ]
                        ]
                        [
                            Models
                            [
                                DrugGPT (Li et al.{,} 2023){,} LiGAN (Masuda et al.{,} 2020){,}  \\ Pocket2Mol (Peng et al.{,} 2022){,} Luo et al. (2021){,} \\ TargetDiff (Guan et al.{,} 2023){,} DiffSBDD (Schneuing et al.{,} 2022), leaf, text width=33em
                            ]
                        ]
                    ]
                    [
                        Conformation \\ 
                        Generation
                        [
                           Datasets 
                           [
                            GEOM-QM9{,} GEOM-Drugs (Axelrod et al.{,} 2022){,} \\ ISO17 (Schutt et al.{,} 2017), leaf, text width=33em, fill=mycolor2
                           ]
                        ]
                        [
                            Metrics 
                            [
                                Coverage (Xu et al.{,} 2021){,} Matching (Xu et al.{,} 2021), leaf, text width=33em, fill=mycolor2
                            ]
                        ]
                        [
                            Models
                            [
                                CVGAE (Mansimov et al.{,} 2019){,} GraphDG (Simm et al.{,} 2019){,} \\ CGCF (Xu et al.{,} 2021){,} GeoMol (Ganea et al.{,} 2021){,} \\ ConfGF (Shi et al.{,} 2021){,} DGSM (Luo et al.{,} 2021){,} \\ GeoDiff (Xu et al.{,} 2022), leaf, text width=33em, fill=mycolor2
                            ]
                        ]
                    ]
                ]
                [
                    Protein
                    [
                        Representation \\ Learning
                        [
                            Datasets
                            [  
                                UniProt (Apweiler et al.{,} 2004){,} ProteinKG (Zhang et al.{,} 2022){,} \\ PDB (Berman et al.{,} 2000){,} AlphaFoldDB (Varadi et al.{,} 2022){,} \\ Pfam (Mistry et al.{,} 2021), leaf, text width=33em, fill=mycolor2
                            ]
                        ]
                        [
                            Tasks
                            [
                                Contact Prediction{,} Fold Classification{,} Stability Prediction{,} PPI, leaf, text width=33em, fill=mycolor2
                            ]
                        ]
                        [
                            Metrics
                            [
                                Accuracy{,} Precision{,} Spearman's $\rho$, leaf, text width=33em, fill=mycolor2
                            ]
                        ]
                        [
                            Models
                            [
                                UniRep (Alley et al.{,} 2019){,} ProtBERT (Elnaggar et al.{,} 2021){,} \\ ESM-1B (Rives et al.{,} 2021){,} MSA Transformer (Rao et al.{,} 2021){,} \\ RSA (Ma et al.{,} 2023){,} OntoProtein (Zhang et al.{,} 2022){,} \\ KeAP (Zhou et al.{,} 2023){,} IEConv (Hermosilla et al.{,} 2020){,} \\ DeepFRI (Gligorijevic et al.{,} 2021){,} GearNET (Zhang et al.{,} 2022), leaf, text width=33em, fill=mycolor2
                            ]
                        ]
                    ]
                    [
                        Structure \\ Prediction 
                        [
                            Datasets
                            [
                                PDB (Berman et al.{,} 2000){,} CASP14 (Kryshtafovych et al.{,} 2021){,} \\ CAMEO (Haas et al.{,} 2018), leaf, text width=33em
                            ]
                        ]
                        [
                            Metrics
                            [
                                RMSD{,} GDT-TS (Zemla et al.{,} 2003){,} \\ TM-score (Zhang et al.{,} 2004){,} lDDT (Mariani et al.{,} 2013), leaf, text width=33em
                            ]
                        ]
                        [
                            Models 
                            [
                                AlphaFold2 (Jumper et al.{,} 2021){,} trRosetta (Du et al.{,} 2021){,} \\ RoseTTAFold (Baek et al.{,} 2021){,} ESMFold (Lin et al.{,} 2023){,} \\ EigenFold (Jing et al.{,} 2023), leaf, text width=33em
                            ]
                        ]
                    ]
                    [
                        \bib{Sequence} \\ \bib{Generation}
                        [   
                            Datasets
                            [
                                PDB (Berman et al.{,} 2000){,} UniRef/UniParc (Apweiler et al.{,} 2004){,} \\ CATH (Sillitoe et al.{,} 2015){,} TS500 (Li et al. 2014), leaf, text width=33em
                            ]
                        ]
                        [   
                            Metrics
                            [
                                AAR{,} RMSD{,} Nonpolar Loss{,} PPL, leaf, text width=33em
                            ]
                        ]
                        [
                            Models
                            [
                            ProteinVAE (Lyu et al.{,} 2023){,} ProT-VAE (Sevgen et al.{,} 2023){,} \\ ProteinGAN (Repecka et al.{,} 2021){,} \\ ProteinSolver (Strokatch et al.{,} 2020){,} \\ PiFold (Gao et al.{,} 2022){,} Anand et al. (2022){,} \\ ABACUS-R (Liu et al.{,} 2022){,} ProRefiner (Zhou et al.{,} 2023){,} \\ GPD (Mu et al.{,} 2024){,} GVP-GNN (Jing et al.{,} 2020){,} \\ ESM-IF1 (Hsu et al.{,} 2022){,} ProteinMPNN (Dauparas et al.{,} 2022),leaf, text width=33em
                            ]
                        ]
                    ]
                    [
                        Backbone Design 
                        [
                            Datasets
                            [
                                PDB (Berman et al.{,} 2000){,} AlphaFoldDB (Varadi et al.{,} 2022){,} \\ SCOP (Murzin et al.{,} 1995){,} SCOPe (Chandonia et al.{,} 2022){,} \\ CATH (Sillitoe et al.{,} 2015), leaf, text width=33em
                            ]
                        ]
                        [
                            Metrics
                            [
                                scTM (Trippe et al.{,} 2022){,} scRMSD{,} AAR{,} PPL{,} RMSD, leaf, text width=33em
                            ]
                        ]
                        [
                            Models
                            [
                                ProtDiff (Trippe et al.{,} 2022){,} \\ FoldingDiff (Wu et al.{,} 2022){,} LatentDiff (Fu et al.{,} 2023){,} \\ Genie (Lin et al.{,} 2023){,} FrameDiff (Yim et al.{,} 2023){,} \\ RFDiffusion (Watson et al.{,} 2023){,} \bib{GPDL (Zhang et al.{,} 2023){,}} \\ GeoPro (Song et al.{,} 2023){,} Protpardelle (Chu et al.{,} 2023){,} \\ ProtSeed (Shi et al.{,} 2022), leaf, text width=33em
                            ]
                        ]
                    ]
                ]
                [
                    Antibody
                    [
                        Representation \\ Learning 
                        [
                            Datasets
                            [
                            OAS (Olsen et al.{,} 2022), leaf, text width=33em, fill=mycolor2
                            ]
                        ]
                        [
                            Models
                            [
                                BERTTransformer (Li et al.{,} 2022){,} AntiBERTy (Ruffolo et al.{,} 2021){,} \\ AntiBERTa (Leem et al.{,} 2022){,} AbLang (Olsen et al.{,} 2022){,} \\ PARA (Gao et al.{,} 2023), leaf, text width = 33em, fill = mycolor2
                            ]
                        ]
                    ]
                    [
                        Structure \\ Prediction
                        [
                            Datasets
                            [
                                SAbDab (Dunbar et al.{,} 2014){,} RAB (Adolf et al.{,} 2018), leaf, text width = 33em, fill = mycolor2
                            ]
                        ]
                        [
                            Metrics
                            [
                                RMSD{,} OCD (Marze et al.{,} 2016), leaf, text width = 33em, fill = mycolor2
                            ]
                        ]
                        [
                            Models 
                            [
                                tFold-Ab (Wu et al.{,} 2022){,} xTrimoABFold (Wang et al.{,} 2022){,} \\ ABodyBuilder2 (Abanades et al.{,} 2023){,} \\ AbLooper (Abanades et al.{,} 2022){,} \\ DeepH3 (Ruffolo et al.{,} 2022){,} SimpleDH3 (Zenkova et al.{,} 2021){,} \\ DeepAb (Ruffolo et al.{,} 2022){,} Igfold (Ruffolo et al.{,} 2023), leaf, text width = 33em, fill = mycolor2
                            ]
                        ]
                        ]
                        [
                        CDR Generation
                        [
                            Datasets
                            [
                                SAbDab (Dunbar et al.{,} 2014){,} RAB (Adolf et al.{,} 2018){,} \\ SKEMPI (Jankauskaite et al.{,} 2019), leaf, text width = 33em, fill = mycolor2
                            ]
                        ]
                        [
                            Tasks
                            [
                                Sequence and Structure Modeling{,} CDR-H3 Generation{,} \\ Affinity Optimization, leaf, text width = 33em, fill = mycolor2
                            ]
                        ]
                        [
                            Metrics
                            [
                                AAR{,} RMSD{,} TM-score (Zhang et al.{,} 2004){,} $\mathbf{\Delta\Delta G}$, leaf, text width = 33em, fill = mycolor2
                            ]
                        ]
                        [
                            Models 
                            [
                                Akbar et al. (2022){,} RefineGNN (Jin et al.{,} 2021){,} \\ MEAN (Kong et al.{,} 2022){,} AntiDesigner (Tan et al.{,} 2023){,} \\ DiffAb (Luo et al.{,} 2022){,} DockGPT (McPartlon et al.{,} 2022){,} \\ HERN (Jin et al.{,} 2022){,} dyMEAN (Kong et al.{,} 2023), leaf, text width = 33em, fill = mycolor2
                            ]
                        ]
                        ]
                    ]
                    [
                    \bib{Peptide}
                    [
                        Misc. Tasks 
                        [
                            Models 
                            [
                                MMCD (Wang et al.{,} 2024){,} PepGB (Lei et al.{,} 2024){,} \\ PepHarmony (Zhang et al.{,} 2024){,} PEFT-SP (Zeng et al.{,} 2023){,} \\ AdaNovo (Xia et al.{,} 2024), leaf, text width = 33em, 
                            ]
                        ]
                    ]
                    ]
                ]
        \end{forest}
    \end{adjustbox}
    \vspace{0.5cm}
    \caption{A structured layout for all terms and papers covered in our survey, including datasets, models, and metrics for each task. Sections contained in the main text are highlighted in blue, while sections expanded upon in the appendix are highlighted in purple.}
    \label{fig:tree}
       
\end{figure*}